\documentclass[11pt,a4paper]{article}

\usepackage[utf8]{inputenc}
\usepackage[T1]{fontenc}
\usepackage[english]{babel}
\usepackage{geometry}
\usepackage{hyperref}
\usepackage{booktabs}
\usepackage{tabularx}
\usepackage{longtable}
\usepackage{array}
\usepackage{amsmath}
\usepackage{amssymb}

\usepackage{newtxtext}
\usepackage{newtxmath}
\usepackage{graphicx}
\usepackage{xcolor}
\usepackage{enumitem}
\usepackage{titlesec}
\usepackage{abstract}
\usepackage{authblk}

\usepackage{setspace}
\usepackage{fancyhdr}
\usepackage{natbib}
\usepackage{tcolorbox}
\usepackage{listings}
\usepackage{caption}

\definecolor{darkviolet}{RGB}{74,25,66}
\definecolor{darkgray}{RGB}{80,80,80}
\definecolor{lightgray}{RGB}{245,245,245}
\definecolor{criticalred}{RGB}{180,30,30}
\definecolor{importantyellow}{RGB}{160,120,0}
\definecolor{suggestiongreen}{RGB}{30,100,50}

\hypersetup{
  colorlinks=true,
  linkcolor=darkviolet,
  citecolor=darkviolet,
  urlcolor=darkviolet,
  pdftitle={IACDM: Interactive Adversarial Convergence Development Methodology},
  pdfauthor={Jasmine Moreira},
  pdfsubject={AI-Assisted Software Development Methodology}
}

\titleformat{\section}
  {\large\bfseries\color{darkviolet}}
  {\thesection.}{0.5em}{}

\titleformat{\subsection}
  {\normalsize\bfseries}
  {\thesubsection.}{0.5em}{}

\tcbuselibrary{skins}
\newtcolorbox{thesis}{
  colback=lightgray,
  colframe=darkviolet,
  fonttitle=\bfseries,
  left=6pt, right=6pt, top=4pt, bottom=4pt
}

\newtcolorbox{finding}[1]{
  colback=lightgray,
  colframe=darkgray,
  title={#1},
  fonttitle=\small\bfseries,
  left=6pt, right=6pt, top=4pt, bottom=4pt
}

\title{
  \vspace{-0.5cm}
  {\large Technical Foundation Document}\\[0.5em]
  {\LARGE\bfseries IACDM: Interactive Adversarial Convergence\\Development Methodology}\\[0.3em]
  {\large A Structured Framework for AI-Assisted Software Development}
}

\author{Jasmine Moreira}
\date{2025--2026}

\begin{document}

\maketitle
\thispagestyle{fancy}

\begin{abstract}
\noindent
Adoption of AI-assisted development in 2025 --- and the emergence of \emph{vibe coding} \citep{Karpathy2025}, generating complete applications from natural language without verification --- exposed a tool-agnostic failure pattern: experienced developers using frontier models were measurably slower while believing they were faster \citep{METR2025}, and 10.3\% of applications in a production showcase leaked data through misconfigured access policies \citep{Palmer2025}. We argue these failures share a structural cause, the \emph{verification gap}: absent external tool use, no language model can determine whether what it generated is correct. The tool is irrelevant; the process is determinative.

We present IACDM (\emph{Interactive Adversarial Convergence Development Methodology}), an 8-phase framework addressing the verification gap through verification agents external to the generator, operating at discrete gates. In each phase the AI alternates between building artifacts and attacking them through specialized lenses, until the design is robust enough to implement. Three pillars: problem discovery via Hierarchical Semantic Analysis before any technical solution; persistent knowledge management across sessions; and systematic adversarial critique before implementation.

What distinguishes the approach from the review and red-teaming traditions it borrows from is that the gate is enforced by a state machine external to the model rather than by convention: the agent may request advancement, not grant it. When one implementation lost that enforceability, the model began self-reporting compliance and the build was withdrawn --- reported here as a single uncontrolled observation, not a comparison.

One component has been tested. A pre-registered experiment over twelve projects under a single frozen instrument asked whether the nineteen critique lenses are non-redundant, against a prior expectation drawn from replications of Perspective-Based Reading, which found reading perspectives not to differ. Every lens found at least one defect that no other lens found --- its exclusive contribution was never zero --- and this held under four independent judgements of what counts as the same defect, including the most aggressive. Mean overlap across lenses was 11\%, and the pair the method's own analysis plan named as most likely to duplicate shared nothing. Two further results were not planned: the taxonomy withstands testing while the criterion deciding when to apply it does not, and the method creates commitments --- architectural premises, resolutions of critical findings --- that nothing ever re-confronts with the built system.

The method was otherwise developed over innovation projects at INDT, an industrial R\&D institute in Manaus, Brazil, plus a varied evaluation set, built because innovation work is unrepeatable and cannot carry a control arm. \textbf{That evidence establishes applicability, not effectiveness}: no project was run without the method, and the proponents are the evaluators. Claims are marked by evidentiary status, and each unvalidated one is restated as a testable hypothesis (H1--H9). Whether IACDM outperforms direct prompting remains open, and this paper is written to make it refutable.
\end{abstract}

\noindent\textbf{Keywords:} AI-assisted software development, adversarial critique, verification gap, methodology, large language models, software engineering, red teaming, iterative design

\vspace{0.5em}
\noindent\textbf{Repository:} \href{https://github.com/jasminemoreira/Versus}{github.com/jasminemoreira/Versus}

\vspace{0.3em}
\noindent\textbf{Validation data (Section~\ref{sec:ro3}):} \href{https://doi.org/10.5281/zenodo.21908908}{doi:10.5281/zenodo.21908908}
\vspace{0.3em}
\noindent\textbf{VSCode Extensions:} Actively maintained builds for Claude, Qwen, Kimi, and local models served through Ollama, at \href{https://marketplace.visualstudio.com/publishers/JasmineMoreira}{marketplace.visualstudio.com/publishers/JasmineMoreira}. A GitHub Copilot build was discontinued in May 2026; see Section~\ref{sec:impl}.

\hrule
\vspace{1em}

\tableofcontents

\newpage

\section{Motivation: the vibe coding phenomenon and its limits}\label{sec:motivation}

In 2025, adoption of AI-assisted development tools reached critical mass. Platforms such as Lovable, Bolt.new, Replit, Cursor, Claude Code, and ChatGPT enabled anyone --- with or without programming experience --- to generate complete applications from natural language descriptions. The phenomenon, coined \emph{vibe coding} by \citet{Karpathy2025}, transformed public perception of who can build software and at what speed.

The numbers are striking: Lovable reached US\$100M in annual recurring revenue (ARR) in 8 months \citep{Lovable2025}; 25\% of Y Combinator's Winter 2025 batch built startups with over 95\% AI-generated code \citep{YC2025} (blog post; not peer-reviewed); the 2025 Stack Overflow Developer Survey reported that 65\% of developers use AI tools daily or weekly \citep{StackOverflow2025}. The productivity case is compelling and, for rapid prototyping, frequently borne out. However, 2025 empirical data reveals a consistent failure pattern that motivated the development of Iterative Convergence.

\subsection{Empirical evidence: the cost of the verification gap}

\paragraph{The METR study (2025).}
A randomized controlled trial (RCT) by METR --- Model Evaluation \& Threat Research --- recruited 16 experienced open-source developers (with moderate AI experience) to complete 246 tasks in open-source repositories where they had an average of 5 years of prior experience \citep{METR2025}. The study used Cursor Pro with Claude 3.5/3.7 Sonnet as the primary AI tool. A notable three-way divergence emerged between what developers expected, what they felt, and what was measured: initial forecasts predicted a 24\% speedup; post-task self-assessment reported a 20\% speedup; objective measurement revealed a 19\% slowdown \citep{METR2025}. The 39-percentage-point gap between perception and reality is the central empirical motivation for this work.

\paragraph{CVE-2025-48757: Lovable.}
In March 2025, security researcher Matt Palmer developed an automated script to scan applications from Lovable's showcase platform. Of the 1,645 projects analyzed, roughly 10.3\% --- 170 applications --- exposed sensitive data through misconfigured or absent Row Level Security (RLS) policies, totaling 303 vulnerable endpoints leaking names, emails, payment data, API keys, and personal addresses \citep{Palmer2025}. The root cause was structural: Lovable's client-driven architecture delegated security responsibility to application developers, most of whom lacked the expertise to implement correct RLS configurations. A Palantir engineer independently demonstrated active exploitation by extracting sensitive data --- personal debt amounts, home addresses, and API keys --- from multiple applications \citep{Palmer2025}.

\paragraph{Degradation at scale.}
A five-year longitudinal analysis by GitClear across 211 million changed lines of code revealed a structural shift in how AI-assisted code is composed \citep{GitClear2025} (industry report; not peer-reviewed). Between 2021 and 2024, the share of changes attributable to refactoring activity fell from 25\% to under 10\%, while copy-pasted code grew from 8.3\% to 12.3\% --- with 2024 marking the first year in GitClear's data where copy-pasted lines exceeded moved lines. The 2025 Stack Overflow Developer Survey registered the first significant decline in positive sentiment toward AI tools --- from over 70\% in 2023 and 2024 to 60\% in 2025 --- even as adoption rose to 84\% of respondents; only 3\% of developers report high trust in AI output accuracy \citep{StackOverflow2025}.

\subsection{The central observation: the tool is irrelevant}

The critical point is that these problems are not specific to any tool. Lovable uses Claude. Claude Code uses Claude. ChatGPT uses GPT. Cursor uses both. All operate as stochastic generators:
\[
P(\text{token}_n \mid \text{token}_1, \ldots, \text{token}_{n-1})
\]
None verify what they produce internally. Absent external tool use, semantic verification capability is zero in every case --- a model cannot determine whether what it generated is correct without feedback from outside itself.

The METR study confirms this: developers used Cursor Pro with frontier models --- tools considered ``serious'' --- and were still slower. Differences between interfaces (visual, terminal, chat), integrators (Supabase, GitHub), or models (Claude, GPT) are cosmetic from the verification gap perspective. The real distinction is not between tools. It is between two modes of operation (Table~\ref{tab:comparison}):

\begin{table}[ht]
\small
\begin{tabularx}{\textwidth}{>{\raggedright\arraybackslash}X >{\raggedright\arraybackslash}X}
\toprule
\textbf{Without methodology (any tool)} & \textbf{With Iterative Convergence (any tool)} \\
\midrule
\texttt{prompt → GA → artifact → delivery} & \texttt{design → G\textsubscript{0} → critique → G\textsubscript{1} → \ldots → G\textsubscript{n} → delivery} \\[4pt]
Verification gap open & Discrete gates with VA-automatic and VA-human \\[4pt]
Errors detected at \(t = \text{production}\) & Errors detected at \(t = \text{gate}_k\) \\[4pt]
Perception: +20\% $|$ Reality: $-$19\% \citep{METR2025} & Measurable at each gate \\[4pt]
170 vulnerable apps / 1,645 analyzed \citep{Palmer2025} & Security verified by VA before delivery \\
\bottomrule
\end{tabularx}
\caption{Comparison of development modes: direct prompting vs.\ Iterative Convergence.}\label{tab:comparison}
\end{table}
\noindent\footnotesize{GA = Generative Agent (the LLM); VA = Verification Agent (automatic or human); $G_k$ = gate at phase $k$. Formal definitions in Section~\ref{sec:verification}.}\normalsize

\begin{thesis}
\textbf{Core thesis:} The quality of AI-assisted software is determined by the process, not the tool.
\end{thesis}

\subsection{What is borrowed and what is new}\label{sec:contribution}

Most of the machinery in this document is inherited, and it is worth conceding that
plainly before claiming anything. Verification external to the producer is old: code
review, QA sign-off, staged approval, and red teaming all long predate large language
models, and Section~\ref{sec:traditions} maps each borrowing to its source. A reader who concludes that
``put a gate between generation and delivery'' is not a new idea is correct.

The contribution is what changes when the party being gated is a stochastic generator.
Five things do not carry over from the human-process tradition:

\begin{enumerate}[leftmargin=*]
  \item \textbf{The adversary is the same agent that generated.} Perspective-Based Reading \citep{Basili1996} already assigns distinct critical perspectives to reviewers of one artifact, and has controlled-experiment support this work lacks. But it assigns them to \emph{different people}, for whom agreement bias toward the author is not a property of the reading mechanism. Asking a model to critique its own output fails by default, because RLHF-trained models tend to affirm what they are shown \citep{Sharma2023, Perez2023}. Sycophancy is a problem PBR never had to solve, and it is why IACDM's answer cannot be an adversarial persona --- which the same bias softens --- but must be structured lens criteria specifying coverage the model produces regardless of disposition (Section~\ref{sec:lensmodel}).
  \item \textbf{Convergence is a condition on the design, not on the run.} Automated pipelines terminate: turn budget exhausted, task marked done, tests green. IACDM's gate $G_4$ instead asks whether the \emph{artifact} has stopped moving --- structural change below threshold between consecutive versions, with no critical findings outstanding. A run can terminate without converging, and a converged design can still be revised. Stopping conditions bound cost; convergence criteria bound the design's remaining instability, and only the second is a statement about quality.
  \item \textbf{The gate is enforceable against the agent.} In human process a gate is a social convention, waivable by the party it constrains. Here it is a state machine the model does not control, which refuses transitions on unmet criteria (Section~\ref{sec:veto}). This is the sharpest departure and the one most easily missed when the method is described in prose.
  \item \textbf{Discovery targets meaning, not requirements.} Models operate over tokens, not domain semantics. Hierarchical Semantic Analysis and the teach-back inversion (Section~\ref{sec:discovery}), in which the model restates the problem in domain language so the operator can judge whether it matches the intended one, exist to force a semantic anchoring the model does not produce on its own. Requirements engineering never had to solve this, because human analysts already share the referent.
  \item \textbf{The failure catalogue is model-specific.} The antipatterns in Section~\ref{sec:antipatterns} --- validation theater, reference-free implementation, verified-solution bypass, scope presence blindness --- are failure modes of generation-with-plausibility-as-proxy. They have no counterpart in a catalogue written for human teams. Section~\ref{sec:ro3method} traces four of them from the built system back to the commitment they broke.
\end{enumerate}

The synthesis, not any single mechanism, is the claim. Readers who find the individual
ingredients familiar are reading it correctly; the assertion is that this particular
assembly addresses the verification gap, and that assertion is what Section~\ref{sec:limitations} formalizes
as testable rather than asserting as established.

\section{The problem: AI as a speed multiplier without direction}

\subsection{The nature of the generative agent}

An LLM operates as a conditional distribution function, selecting each token based on the probability distribution conditioned on context. This process has three critical properties for software development:

\begin{enumerate}[leftmargin=*]
  \item \textbf{Absence of internal verification.} The model has no mechanism to evaluate the semantic, logical, or functional correctness of what it produces. It does not distinguish correct code from incorrect code --- only more or less probable text given the context \citep{Huang2023}.
  \item \textbf{Plausibility as proxy for correctness.} Output is optimized for statistical plausibility, not correctness. Code that ``looks right'' is favored regardless of its functional correctness \citep{Dziri2023}.
  \item \textbf{Invariance of the generative engine.} The same stochastic mechanism operates regardless of whether the prompt is direct or structured. The conditioning quality changes; semantic verification capability, absent external tool use, remains zero.
\end{enumerate}

These three properties are invariant across interfaces, models, and integration layers. Improving the tool does not close the verification gap; only external verification does.

\subsection{The verification gap}

Let $G(\text{prompt}) \to \text{artifact}$ be the LLM's generative function. For any artifact produced, there exists no internal function $V(\text{artifact}) \to \{\text{correct}, \text{incorrect}\}$ available to the model. This constitutes the verification gap: the space between what is generated and what is verifiable by the generator itself.

The cost manifests in three forms: \emph{context erosion} (code generated without architecture produces monolithic artifacts that degrade subsequent interactions); \emph{accelerated technical debt} (ease of generation incentivizes quick fixes that accumulate coupling and inconsistencies --- Lehman's law of increasing complexity \citep{Lehman1980} operates at multiplied speed); and \emph{validation theater} (developers ask the AI if the design is good; the AI, through RLHF alignment, tends to agree --- and developers, susceptible to processing fluency as a proxy for truth \citep{Kahneman2011}, mistake a confident and well-formed response for a correct one).

\section{Iterative Convergence: formal definition}

IACDM (\emph{Interactive Adversarial Convergence Development Methodology}) is a software development framework that organizes work into 8 phases (0--7), each with entry criteria, expected artifact, and exit gate. Throughout this document, \emph{IACDM} and \emph{Iterative Convergence} refer to the same method: IACDM is the formal acronym used in academic contexts; Iterative Convergence is the operational name used in the project repository and practitioner documentation. The method uses AI as a partner in each phase, alternating between builder and adversary roles, until the design reaches sufficient robustness for implementation.

\subsection{Evidentiary status of claims in this document}\label{sec:status}

This is a technical foundation document: it specifies a method and reports the reasoning
behind its design. Different claims in it rest on very different footings, and conflating
them would be the first failure the method itself is meant to prevent.

Claims whose evidentiary basis is not evident from their context therefore carry one of
four markers. The convention is applied where the basis could be mistaken --- above all
wherever a design decision or a field observation might otherwise read as a result --- and
not to expository prose, definitions, or statements whose warrant is the citation attached
to them. An unmarked sentence is not an unmarked claim; where a marker would change how a
sentence should be read, it is there.

\begin{itemize}[leftmargin=*]
  \item \textbf{[L] Literature-derived.} Follows from cited published work. The citation is the warrant; disagreement is with the source.
  \item \textbf{[D] Design decision.} A choice made in constructing the method, with its rationale stated. Not a finding: a different designer could choose otherwise. The claim is that the rationale is coherent, not that the choice is optimal.
  \item \textbf{[O] Field observation.} Observed across the two-tier project corpus of Section~\ref{sec:emplimits} --- institutional innovation projects plus the evaluation set. \emph{These observations were not instrumented, timed, or tabulated} --- they are the author's and practitioners' recollection of recurring patterns, subject to every bias that entails. They motivate hypotheses; they do not test them.
  \item \textbf{[H] Untested hypothesis.} Stated so it can be refuted. Section~\ref{sec:limitations} collects these into a validation roadmap.
\end{itemize}

A fifth class exists and deliberately has no marker. Some quantities in this document ---
per-phase durations, iteration counts, the distribution of projects across models --- were
\emph{extracted from machine-written project records} rather than recalled. They are
neither [O], since nothing about them is recollection, nor measurements in the sense a
controlled study would produce, since the records were a by-product of using the method
rather than an instrument built to test it. A marker would compress that into three
letters and mislead. Wherever such a quantity appears, its basis and its limits are stated
where it is used, at length (Section~\ref{sec:record}).

The distinction that matters most is \textbf{[O] versus [H]}. An observation reported by
the proponents of a method, without instrumentation or a control condition, is not weak
evidence for that method's effectiveness --- it is not evidence of effectiveness at all.
It is a reason to formulate a hypothesis.

One consequence of applying this convention is visible throughout: several claims earlier
versions of this document stated plainly are now marked, qualified, or reported as
refuted. The document says so where it happens rather than quietly revising, because a
method whose subject is the difference between assertion and evidence is poorly placed to
erase its own record of having confused them.

\begin{thesis}
\textbf{Thesis 1 [D]:} Software quality in AI-assisted development is determined by the quality of design preceding code generation, not by the model's generative capability.\\[0.3em]
\textbf{Thesis 2 [D]:} AI is most productive on granular work units with explicit scope, documented assumptions, and well-defined interfaces, maximizing the useful-information/context-consumed ratio.\\[0.3em]
\textbf{Thesis 3 [D]:} Deep problem understanding is a precondition for any technical decision. Understanding the wrong problem is orders of magnitude more expensive than understanding it slowly.
\end{thesis}

\paragraph{Provenance of the three theses.}
They are design premises, not empirical findings --- which is why each carries [D]. Each
was adopted because an established result makes it plausible, and each is stated here in
the strongest form the method actually requires. T1 restates, for the AI-assisted setting,
Brooks' separation between essential and accidental difficulty \citep{Brooks1987}: if
conceptual construction is where the difficulty lives, then the design preceding
generation dominates. T2 follows from the documented degradation of retrieval over long
contexts \citep{Liu2023}: it is an engineering response to a measured property of current
models, and it would weaken if that property did. T3 is the design-phase reading of
Boehm's cost-of-change curves \citep{Boehm1986}; the ``orders of magnitude'' is Boehm's
order-of-magnitude-per-phase claim carried to the discovery phase, not an independently
measured multiplier of our own.

None of the three has been tested \emph{as stated} in an AI-assisted setting. Their
testable forms are H1--H3 in Section~\ref{sec:limitations}.

\subsection{Overview of the 8 phases}\label{sec:phases}

\begin{longtable}{clp{4.5cm}p{4.5cm}}
\toprule
\textbf{\#} & \textbf{Phase} & \textbf{Purpose} & \textbf{Primary artifact} \\
\midrule
\endhead
0 & Problem Discovery & Understand domain and problem before any technical solution & Validated document (score $\geq$90/100) + specs/ populated \\
1 & Architecture & Define modules, interfaces, and technical contracts & Architecture V1 ($\sim$2k tokens; author estimate [D]) \\
2 & Adversarial Critique & Attack architecture with up to 19 specialized lenses (7 universal + up to 12 conditional) & Coverage matrix (modules $\times$ lenses) \\
3 & Simplification & Simplify without introducing bugs; generate V(N+1) & Simplified architecture \\
4 & Convergence Gate & Validate convergence with rigorous criteria & Final approved architecture \\
5 & Code Implementation & Implement without regression; one module at a time & Functional code \\
6 & Tests & Test against interfaces; mandatory manual testing & 100\% tests passing \\
7 & Post-Review & Systematic learning; methodology evolution & Lessons + method improvement proposal \\
\bottomrule
\caption{The 8 phases of IACDM}\label{tab:phases}
\end{longtable}

Table~\ref{tab:phases} summarizes all phases. Phases 2--3 form an iterative loop that repeats until convergence.

\paragraph{How often the loop actually runs --- a corrected claim.}
Earlier statements of this method held that small projects run 1--2 critique iterations, medium ones 2--3, and large or poorly understood ones 3--5. That was recollection, never tabulated. The persisted transition logs (Section~\ref{sec:record}) record it exactly, which made the estimate checkable; checking it refutes it.

The validation batch of Section~\ref{sec:ro3} gives the only figure produced under a
controlled instrument: across its twelve projects the loop ran 28 times in total ---
nine projects with two iterations, two with three, one with four. That is a narrower and
lower range than the estimate claimed, on projects deliberately sized at 8--12 modules and a
single session.

It does not settle the estimate, because the batch holds project size roughly constant and
the estimate was about how iteration count varies \emph{with} size. What can be said is that
nothing in a controlled corpus reached the upper end, and that the figure remains
\textbf{[O]} --- recollection, unverified for the range it actually claims.

The stopping criterion belongs to the human operator: structural change below 15\% between consecutive versions and zero critical findings.

\paragraph{Provenance of the convergence thresholds.}
Both numbers are \textbf{[D]}, and neither was fitted to data.

The \emph{15\% structural change} threshold operationalises a decision that is qualitative in nature: whether the design is still moving. It was set at the point where, in the author's judgement, successive versions had stopped differing in structure and started differing in wording --- and it is deliberately paired with the second condition, zero criticals, precisely because a change-magnitude number alone is a poor proxy for convergence. A design can churn below 15\% and still be broken; the critical-findings condition, not the percentage, is what carries the safety argument. The percentage exists to stop \emph{perfectionist} iteration (AP8), not to certify correctness.

The \emph{score $\geq$90} gate for Phase~0 is a satisficing threshold, and it is not a lone cut but the top band of a three-band scale: below 70, the problem is far enough from understood that several more iterations are expected; 70--89 calls for one more, targeted at the lowest-scoring criteria; 90 or above advances. The operative instruction attached to it in the tooling is a ceiling, not a floor --- \emph{90 is sufficient; do not aim for 100} --- and it is anchored explicitly to AP8, convergence perfectionism. The threshold exists to stop iteration as much as to gate it.

Two properties of the number follow from the weights (Section~\ref{sec:discoveryloop}) rather than motivating it, but are worth stating because they bound what the gate can let through. Permitting at most 10 missing points makes it arithmetically impossible to advance while failing either 15-point criterion --- complete use cases, or resolved ambiguities --- while still tolerating one 10-point gap or both 5-point gaps. So the gate is strict exactly where the weights say the risk concentrates, and forgiving elsewhere.

The weights are the more consequential calibration, and they are the less defensible one: nothing establishes that use cases and ambiguities deserve 15 while success criteria and validated assumptions deserve 10. That ordering is judgement. Changing it would change what the gate enforces far more than moving the threshold by five points would.

\subsection{Activation criteria}

The methodology does not apply to every AI interaction. Activation is based on observable signals of complexity, not LOC estimation --- since estimating lines of code before understanding the problem is precisely the kind of implicit assumption the method criticizes.

\begin{thesis}
\textbf{Golden rule:} If you cannot confidently assert that the project is simple, it is not simple.
\end{thesis}

\textbf{[D]} The rule is justified by an asymmetry of \emph{orders of magnitude}, not by a measured ratio. Applying the methodology to a project that turns out simple costs an abbreviated Phase~0 --- on the order of tens of minutes. Not applying it to a project that turns out complex costs rework measured in hours or days, because the defect is conceptual and surfaces late \citep{Boehm1986}. The rule follows from the asymmetry being large and one-sided, and would survive either quantity being off by a factor of two.

\textbf{[O]} The working figure for an abbreviated Phase~0 has been roughly 20 minutes. It is recollection, and it remains so: an attempt to check it against persisted records was withdrawn along with the rest of that analysis (Section~\ref{sec:record}), and the validation batch of Section~\ref{sec:ro3} was designed to answer a different question and does not settle this one. Both halves of the asymmetry are therefore untested --- the cost, and the claim that it is small relative to the rework it averts. That is H4 (Section~\ref{sec:hypotheses}).

\subsection{Relationship to existing methods}\label{sec:traditions}

IACDM integrates principles from diverse software engineering traditions (Table~\ref{tab:traditions}):

\begin{table}[ht]
\small
\begin{tabularx}{\textwidth}{>{\raggedright\arraybackslash}p{3.2cm} >{\raggedright\arraybackslash}X >{\raggedright\arraybackslash}X}
\toprule
\textbf{Tradition / Method} & \textbf{Incorporated principle} & \textbf{Adaptation in IACDM} \\
\midrule
Spiral Model \citep{Boehm1986} & Iterative cycles with risk analysis & Risk evaluated as design fragility; AI is risk analysis partner \\
Design by Contract \citep{Meyer1992} & Pre/postconditions and invariants & Explicit assumptions and negative scope as informal contracts \\
Falsificationism \citep{Popper1959} & Knowledge advances through refutation & Adversarial critique seeks to refute the design, not validate it \\
Red Teaming (NSA/DARPA) & Dedicated team to attack the system & AI assumes red team role under adversarial prompt \\
Perspective-Based Reading \citep{Basili1996} & Reviewers read one document from assigned perspectives, under procedural scenarios & Lenses are the same idea applied to a single generator rather than to distinct reviewers (Section~\ref{sec:lensmodel}) \\
ATAM \citep{Kazman2000} & Scenario-based architectural trade-offs & Each critique cycle evaluates trade-offs, forces explicit decisions \\
KISS/YAGNI \citep{Beck1999} & Simplicity as value & Each iteration must simplify, not complexify \\
TDD \citep{Beck2003} & Tests drive design & Adversarial critique as ``design testing'' before code \\
ADR \citep{Nygard2011} & Decision documentation with context & Documented assumptions serve an equivalent role \\
\bottomrule
\end{tabularx}
\caption{IACDM principles from established software engineering traditions.}\label{tab:traditions}
\end{table}

\section{Problem discovery and knowledge management}\label{sec:discovery}

Phase 0 (Problem Discovery) is the most distinctive contribution of IACDM relative to traditional AI-assisted development methods. It formalizes a central observation: understanding the wrong problem is orders of magnitude more expensive than understanding it slowly --- an estimate consistent with Boehm's cost-of-change curves \citep{Boehm1986}, though the exact multiplier depends on domain and detection phase.

\subsection{The iterative discovery loop}\label{sec:discoveryloop}

Phase 0 operates as an iterative convergence cycle over problem understanding (not solution understanding), with four steps per iteration: (1) \emph{Collection} --- structured AI questions about domain, use cases, vocabulary, constraints, and negative scope; (2) \emph{Synthesis (teach-back)} --- the AI explains back what it understood in domain language; (3) \emph{Validation (confrontation)} --- the operator corrects or points out gaps; and (4) \emph{Convergence assessment} --- score 0--100 based on 10 weighted criteria, exit gate at score $\geq$90 + explicit operator confirmation.

\textbf{[D]} The 10 criteria and their weights are (see §3.2 for how the 90 threshold follows from them): (1) problem clarity --- problem stated without ambiguity (10 pts); (2) complete use cases --- top 3--5 use cases with actors, goals, and flows (15 pts); (3) defined vocabulary --- domain terms agreed and consistent (10 pts); (4) resolved ambiguities --- all identified ambiguities resolved or explicitly accepted (15 pts); (5) explicit out-of-scope --- negative scope declared (10 pts); (6) measurable success criteria --- acceptance criteria verifiable (10 pts); (7) validated assumptions --- implicit assumptions surfaced and confirmed (10 pts); (8) research and \texttt{specs/} populated --- relevant technical and scientific references deposited (5 pts); (9) operator confirmed --- teach-back accepted without major correction (10 pts); (10) AI confident --- AI can answer ``why?'' for every decision without hedging (5 pts). Operator confirmation appears twice by design: as criterion~(9), worth 10 points, and again as a required exit criterion the gate checks independently of the score. A project can therefore reach 90 on the other criteria and still be held at the gate. The weights are subject to revision as validation data accumulates.

\textbf{[D]} The teach-back inverts the direction of agreement bias: instead of the AI confirming it understood --- which it will always affirm --- the human evaluates whether the AI's explanation is correct. \textbf{[L]} In the requirements elicitation literature, the underlying problem --- surfacing tacit knowledge and ambiguity through structured analyst-stakeholder dialogue --- is documented by \citet{Ferrari2016tacit} and \citet{Bano2019}; teach-back as used here operationalizes the same verification intent through AI-mediated paraphrase.

\paragraph{What ``correct'' means here, and how far the check reaches.}
The word carries less weight than it appears to. The operator is not judging truth, and not
checking whether the model reproduced the right words. The test is whether the paraphrase,
expressed in domain language, \emph{matches the referent the operator intended} --- whether
the model's account of the problem would lead a competent reader to the same problem. That
is a judgement of correspondence to intent, and it is irreducibly subjective. No score is
attached to it: criterion~(9) records only that the operator accepted the paraphrase
without major correction, which is a record of an act, not a measure of its quality.

Two limits follow, and both are load-bearing.

First, teach-back cannot detect a misunderstanding the operator shares. If the
operator's own model of the domain is wrong, a paraphrase faithful to that model passes.
The mechanism inverts agreement bias between human and AI; it does nothing about error
common to both. This is the failure mode most likely to survive Phase~0 intact, and
nothing in the method catches it --- external domain expertise would, and IACDM does not
require any.

Second, the operator is exactly the reader the METR result warns about
\citep{METR2025}: a fluent, confident, well-formed paraphrase is precisely what a
plausibility-optimised generator produces, and processing fluency is a known proxy for
perceived truth \citep{Kahneman2011}. The teach-back asks a human to distinguish
\emph{understood} from \emph{fluently restated}, which is the discrimination that study
found experienced developers failing to make. Limitation L4
(Section~\ref{sec:limitations}) is not a footnote to this mechanism; it is a condition on
whether the mechanism works at all.

\textbf{[H]} The claim that teach-back surfaces more misunderstanding than direct
confirmation is untested. It is the most tractable experiment this document proposes
(RO1), it does not require the rest of the methodology, and it would be informative
whichever way it came out.

\subsection{Hierarchical Semantic Analysis (HSA)}

Each word in the name carries a design consequence.

\paragraph{Analysis, not elicitation.} Classical requirements engineering uses the term \emph{elicitation} \citep{Nuseibeh2000} --- a word that presupposes requirements exist ready-formed in the domain, waiting to be extracted. HSA uses \emph{analysis} because the object of the activity is not pre-existing: the understanding of the problem must be \emph{constructed}. This distinction is not terminological. It is what justifies the iterative loop of Phase~0: if requirements could be elicited, a single pass would suffice. Because understanding must be built, decomposition, gap detection, and recomposition are necessary steps, not inefficiencies. The AI does not discover what the operator already knows --- it helps the operator discover what neither party had yet articulated.

\paragraph{Semantic, not structural.} The object of the analysis is not the structure of the problem, nor the expected functionality --- it is the \emph{meaning of the concepts in the domain}. This distinction matters because LLMs operate over tokens, not meanings: a model assigns statistical weight to sequences of text, but has no mechanism to verify whether its use of a term matches the domain-specific meaning the operator intends. HSA is the protocol that forces semantic anchoring the model does not produce automatically. The consequence is visible in the first level of the hierarchy: establishing vocabulary is not a warm-up exercise --- it is the foundational act that determines whether all subsequent understanding is anchored or ambiguous.

The teach-back inversion (Section~\ref{sec:phases}) is intrinsically a semantic test: the operator is not verifying whether the AI repeated the correct words, but whether it captured the correct \emph{meaning}. A notable side effect is that discovering the AI did not capture the meaning frequently reveals that the operator had not yet articulated it with precision. The teach-back thus serves a dual function --- it surfaces misunderstanding in the model and forces the operator to resolve ambiguity in their own mental model.

\paragraph{Hierarchical, not sequential.} A list of steps is sequential; HSA is hierarchical because the levels are related by \emph{ontological foundation}: each level is presupposed by the one above it. It is not possible to understand Processes without having stabilized the Elements that participate in them. It is not possible to specify the Product without having understood the Problem it resolves. This foundational relationship produces two properties absent from flat structures. First, \emph{epistemic restriction}: the operator should not advance to level $N+1$ without convergence at level $N$, because $N$ is the foundation on which $N+1$ rests. Second, \emph{forced retroaction}: when exploration at a deeper level reveals that a concept at a shallower level was imprecisely understood, revision is required upward. This is not a failure of the method --- it is the method functioning as designed.

The hierarchy is simultaneously a hierarchy of \emph{abstraction} (from the general to the specific, top-down) and a hierarchy of \emph{ontological dependency} (each level depends on those below it, bottom-up). These two directions are not the same thing, and both are necessary to understand why the sequence cannot be inverted or collapsed without losing information.

HSA structures exploration into five levels, each building on the previous, imposing epistemic order on the discovery process:

\begin{center}
\begin{tabular}{clp{5cm}}
\toprule
\textbf{Level} & \textbf{Focus} & \textbf{What is sought} \\
\midrule
1. Domain & Universe where the problem exists & Vocabulary, theoretical field, state of the art \\
2. Problem & What needs to be solved & 5W1H analysis \\
3. Elements & Parts that compose the problem & Components, entities, constraints \\
4. Processes & How the parts relate & Flows, dependencies, feedback loops \\
5. Product & What is expected at the end & Deliverables, acceptance criteria, negative scope \\
\bottomrule
\end{tabular}
\end{center}
\noindent\footnotesize{5W1H: Who, What, When, Where, Why, How --- a structured questioning framework for systematic problem decomposition.}\normalsize

\subsection{The knowledge repository (specs/)}

The \texttt{specs/} directory is a persistent structured repository that functions as external project memory, addressing the finite context window of AI and the finite working memory of the human operator simultaneously.

Three properties make it methodologically significant. First, \emph{traceability}: every technical decision traces to literature, and no algorithm is implemented without a bibliographic reference — this prevents the AI from substituting invented parameters for verified ones. Second, \emph{inter-session persistence}: an operator resuming a project after weeks finds accumulated context intact, and the AI queries \texttt{specs/} before asking questions that were already answered in prior sessions. Third, \emph{inter-version transferability}: version 2.0 starts from version 1.0's \texttt{specs/} as a baseline, preserving design rationale across rewrites and preventing regression in understanding.

\section{Architecture before development}

If the essential difficulty of software lies in conceptual construction \citep{Brooks1987} --- and AI tools, by embodying well-established implementation idioms and architectural patterns, primarily address accidental difficulty --- then prior architectural investment is the only way to capture AI's productive contribution. This is an extrapolation of Brooks' framework: the essential difficulty (understanding the right problem, designing coherent abstractions) remains irreducibly human; the accidental difficulty (translating those abstractions into working code) is where AI contributes most reliably. Without prior architecture, AI accelerates production of artifacts that will need to be discarded.

Boehm's foundational research on software cost estimation established that the cost of defect correction grows by an order of magnitude for each phase of delay in discovery \citep{Boehm1986}. IACDM anticipates discovery of conceptual defects to the design phase, when correction cost is orders of magnitude lower than implementation.

Architecture in this context answers four questions: \emph{decomposition} (modules, responsibilities, boundaries --- principle of separation of concerns \citep{Dijkstra1974} and Single Responsibility Principle \citep{Martin2003}); \emph{interfaces} (communication contracts between modules); \emph{assumptions} (what the system takes as true --- Leveson's STAMP framework identifies flawed or outdated mental models of system state as a primary driver of accidents in complex systems \citep{Leveson2011}; undeclared assumptions in software design play an analogous role); and \emph{negative scope} (what the system deliberately does not do --- explicit exclusion prevents scope creep from entering through silence, and gives the AI a boundary it cannot cross without operator approval).

\section{Granularization: maximizing AI context efficiency}\label{sec:granular}

The most important operational constraint of current LLMs is not generation capability but the finite context window. Retrieval performance degrades for information positioned in the middle of long contexts --- a phenomenon documented as \emph{lost in the middle} \citep{Liu2023} --- with broader attention quality implications as context volume grows.

The context efficiency relationship is expressed as a \emph{conceptual definition} (not an empirically computable metric in its current form): $E = I_0/C$, where $E$ = context efficiency, $I_0$ = tokens carrying information directly relevant to the current task, and $C$ = total tokens consumed in the interaction. The ratio is not directly measurable without a relevance oracle, but it captures the intuition: in a monolithic system, $E \to 0$ as $C$ grows unboundedly while $I_0$ (the current module's interface and logic) remains constant. In a granularized system, $C$ is bounded per session and $I_0$ remains a large fraction of it. This definition is used qualitatively throughout this paper; controlled measurement is deferred to the validation roadmap (L3, Section~\ref{sec:limitations}).

Decomposition follows high internal cohesion and low external coupling \citep{Constantine1979}: each module must be comprehensible by the AI in a single interaction, and interfaces --- in the sense of Meyer's abstract module boundaries \citep{Meyer1988} --- function as compact summaries of what adjacent modules expose, allowing each session to remain focused and contextually dense.

\section{Adversarial critique through specialized lenses}

\subsection{Why conventional validation fails}\label{sec:whyfails}

LLMs trained with RLHF exhibit a documented sycophantic bias: they tend to agree with, validate, or soften criticism of whatever the user presents \citep{Sharma2023, Perez2023}. When asked ``is this design good?'', the model tends to agree --- what \citet{Argyris1977} called \emph{single-loop learning}: accepting the problem framing and optimizing within it without questioning the framing itself. IACDM requires \emph{double-loop learning}: questioning not only the solution but the assumptions that sustain it.

Generic critique also fails: it produces superficial observations without systematic coverage of quality dimensions. The same LLM operates all rounds --- without operationally distinct criteria, diversification is illusory. The mechanism that operationalizes adversarial posture is not the prompt framing alone --- which the same sycophantic bias could still soften --- but the structured lens criteria: explicit, non-negotiable coverage requirements that force the model to seek failure modes it would otherwise omit.

\subsection{The specialized lens model}\label{sec:lensmodel}

Phase 2 applies up to 19 specialized lenses organized in three categories. The selection principle governing all categories: \emph{a lens is legitimate only if removing it exposes a class of failure that no other lens detects.}

\paragraph{The antecedent, and what its evidence says about this design.}
\textbf{[L]} Assigning distinct reading perspectives to reviewers of one document is not new.
Perspective-Based Reading formalised it in the 1990s: reviewers work through procedural
scenarios as user, designer, or tester, on the assumption that different perspectives find
different defects and that combining them covers more than the same reading effort under
one perspective \citep{Basili1996}. Controlled experiments with professional developers
found PBR teams achieved significantly better coverage than the reviewers' own usual
technique, and later comparisons found it more effective and cheaper per defect than
checklist-based reading. This document should have cited that literature from the outset;
it is the closest antecedent the lens model has.

Two consequences follow, and they point in opposite directions.

The first supports Section~\ref{sec:whyfails}. The claim that \emph{structured} critique outperforms
unstructured critique has independent empirical backing that IACDM does not possess about
itself. The argument that generic ``attack this design'' produces shallow coverage, and
that explicit per-lens criteria are what force the model past it, is not a conjecture
original to this work --- it is the finding PBR replicated.

The second cuts against the selection principle stated just above. Subsequent replications
asked whether the perspectives are really distinct, and found no significant
difference between them in defect detection rate or in defect coverage, concluding that
combining perspectives may not yield higher coverage than reading under a single one
\citep{Regnell2000}. That is direct counter-evidence to the orthogonality assumption on
which the entire lens taxonomy rests. If perspectives were not separable for human
reviewers reading requirements, there is no prior reason to expect 19 of them to be
separable for a language model critiquing an architecture.

\textbf{[H]} The honest position is therefore narrower than the one this document
previously implied. Structured lenses are warranted; \emph{the specific taxonomy of 19,
and the orthogonality criterion that justifies its size, are not}. Whether lens
independence holds here is RO3 (Section~\ref{sec:limitations}) --- which is now not an
open question but a replication of one already answered negatively in the adjacent
literature. A null result would not invalidate the critique phase; it would mean the lens
set is redundant and should be much smaller.

\paragraph{Universal lenses (7 --- always applied).} Applied to every project regardless of domain or context. Omitting any universal lens leaves a failure class unconditionally unexamined (Table~\ref{tab:universal-lenses}):

\begin{table}[ht]
\small
\begin{tabularx}{\textwidth}{>{\raggedright\arraybackslash}p{2.4cm} >{\raggedright\arraybackslash}X >{\raggedright\arraybackslash}p{4cm}}
\toprule
\textbf{Lens} & \textbf{Central question} & \textbf{Exclusive failure class} \\
\midrule
Assumptions & What does this design assume without declaring? & Failures from flawed/outdated system models \citep{Leveson2011} \\
Architectural & Can each module be replaced, removed, or tested in isolation? & Hidden coupling, circular dependencies \\
Implementability & Can I code this module in one session with available context? & Incomplete specs, insufficient granularization \\
Scientific & Does every value, formula, and algorithm have a verifiable reference? & Invented parameters, plausibility-based logic \\
Security & How would an attacker exploit this surface with minimum effort? & Unanalyzed attack surface \\
Performance & Where are the bottlenecks and what is the asymptotic behavior? & Hidden bottlenecks, scale degradation \\
Regulatory & Does every regulatory requirement trace to a module? & Regulatory non-compliance \\
\bottomrule
\end{tabularx}
\caption{Universal lenses (always applied).}\label{tab:universal-lenses}
\end{table}

\paragraph{Situational lenses (8 --- activated by project context).} Applied when the project exhibits specific structural characteristics. Each lens is declared active or inactive at Phase~1 with explicit rationale (Table~\ref{tab:situational-lenses}):

\begin{table}[ht]
\small
\begin{tabularx}{\textwidth}{>{\raggedright\arraybackslash}p{2.4cm} >{\raggedright\arraybackslash}p{3.2cm} >{\raggedright\arraybackslash}X}
\toprule
\textbf{Lens} & \textbf{Activate when} & \textbf{Exclusive failure class} \\
\midrule
Resilience & External dependencies (APIs, DBs, queues) & Cascading failures, retry storms \\
UI/UX & Any surface a \emph{person} operates --- including a CLI or operational tooling, not only graphical end-user interfaces & Confusing flows, dead-end states, missing feedback, accessibility failures \\
Migration / Coexistence & Replacing or modifying existing production system & Data loss, regression vs.\ legacy, impossible rollback \\
Sustainability / Proportionality & The system decides, allocates, or consumes a resource whose cost grows with use --- e.g.\ but not only AI/ML, high-volume data, elastic infrastructure & GPU where CPU suffices, heavy model for a simple task, infinite data retention \\
Ethical / Human Impact & Automated decisions affecting people & Algorithmic bias, absence of human recourse \\
Process / Workflow & Multi-actor flows, state machines, business processes & Orphaned states, missing actors, happy-path bias \\
Governance / Accountability & The system records or decides something someone must later audit, attribute, or answer for. A \emph{system} property: a single-operator project activates it when the system has it & No data ownership, no accountability, shadow data flows \\
Observability / Operability & Production systems with operational requirements & Opaque systems that cannot be diagnosed in production \\
\bottomrule
\end{tabularx}
\caption{Situational lenses (activated by project context).}\label{tab:situational-lenses}
\end{table}

\paragraph{Three triggers in this table were rewritten mid-experiment, and why.}
The triggers for UI/UX, Sustainability, and Governance shown above are \emph{not} the ones
this method carried until August 2026. They were rewritten because the original wording
made three lenses unusable: across the first seven projects of the validation batch
(Section~\ref{sec:ro3}), Governance and Sustainability activated zero times, and
no amount of rewriting project briefs changed it.

The diagnosis is the finding, not the fix. A trigger written as an \emph{organisational}
condition --- ``multiple teams, data domains, compliance'' --- cannot be exercised by a
central question that asks about a \emph{system} property --- ``is every action
attributable?''. The two describe different kinds of thing, so a reader satisfying one
routinely fails the other. UI/UX had the same defect in milder form: ``user-facing
interface'' left undecided whether a command-line tool counts, and readers split
consistently on it.

Two consequences must be stated plainly. First, \textbf{the orthogonality result reported in
Section~\ref{sec:ro3} measured the corrected taxonomy, not the one previously published}.
Second, this is the clearest instance of the second result of that experiment: the lens
\emph{set} withstood testing while the criterion deciding \emph{when to apply it} did not. The triggers tabulated above are those of the instrument version the experiment ran under (Section~\ref{sec:ro3}). The tooling continues to develop, and a reader installing a later build may find them revised again; this document reports what was measured, not the current state of an evolving implementation.

\paragraph{Domain Transfer Lenses (4 --- activated by signal detection).} A second-order critique layer that applies processes from domains external to software engineering. Unlike universal and situational lenses --- which criticize the design using software vocabulary --- Domain Transfer Lenses import invariants from domains with historically documented conceptual loans to computer science, exposing failure classes that software-internal lenses cannot name because they lack the corresponding concept. Each lens is activated when a specific signal is detected in the problem statement or architecture (Table~\ref{tab:domain-lenses}):

\begin{table}[ht]
\small
\begin{tabularx}{\textwidth}{>{\raggedright\arraybackslash}p{2.4cm} >{\raggedright\arraybackslash}X >{\raggedright\arraybackslash}p{3cm}}
\toprule
\textbf{Lens} & \textbf{Central question / Exclusive failure class} & \textbf{Activation signal} \\
\midrule
Control Engineering & Where does the system generate an error signal and correct it? Risk of oscillation or state drift? \emph{Failure class: systems that react to events but do not regulate state --- oscillation, drift, runaway feedback.} & State synchronization, runtime configuration affecting behavior, feedback-driven systems \\
Game Theory & Do system actors have aligned incentives? Where does the design assume cooperation and may encounter strategic defection? \emph{Failure class: architectures that work under cooperation assumptions but collapse under adversarial or strategic behavior.} & Multiple independent actors, public API, marketplace or platform design \\
Linguistics / Grammar & Is the interface contract unambiguous? Can two correct implementations of the same contract produce incompatible behaviors? \emph{Failure class: protocol ambiguity --- two correct implementations that are mutually incompatible.} & Inter-component communication, message formats, interface contracts between independent teams \\
Mechanical Engineering & Where are the tolerances? Does the system work only at exact specification or does it tolerate variation? \emph{Failure class: rigid coupling disguised as tolerance --- failure from small deviations in dependency versions, environment, or load.} & Module maintenance, long-lived systems, technical debt accumulation \\
\bottomrule
\end{tabularx}
\caption{Domain Transfer Lenses (activated by signal detection).}\label{tab:domain-lenses}
\end{table}

The corpus of transferable domains is grounded in a two-part criterion: (1) the domain must have transformational and logical processes that are formally documented, and (2) the conceptual loan to computer science must have historical precedent. The four lenses in the initial corpus satisfy both criteria: control engineering concepts underpin reactive systems and PID controllers; game theory informs distributed consensus and protocol design; linguistic frameworks shaped API and DSL design; mechanical tolerances informed coupling and cohesion theory. The corpus is extensible by the same criterion.

\subsection{The critique--response asymmetry}

An important property of the lens model is the asymmetry between critique and response: multiple agents to attack, unified agent to respond, human operator to arbitrate trade-offs. This asymmetry reflects ATAM practice \citep{Kazman2000}, where different evaluators attack by distinct quality attributes, but the architect resolves the trade-offs. The three-category lens architecture --- universal, situational, and domain transfer --- amplifies this asymmetry: each category introduces a qualitatively distinct attack vector, so the unified response in Phase~3 must integrate software-internal findings with cross-domain structural critiques.

\subsection{The coverage matrix}\label{sec:matrix}

The primary artifact of Phase 2 is the coverage matrix: a two-dimensional map (modules $\times$ lenses) showing finding distribution by severity. The gate criterion verifies \emph{completeness of critique}, not absence of findings: \textcolor{criticalred}{critical} findings are recorded and carried into Phase~3 for resolution --- they do not block advancement, because Phase~3 exists precisely to resolve them. Each \textcolor{importantyellow}{important} finding requires an explicit decision: resolve in Phase~3, or accept risk with documented justification. \textcolor{suggestiongreen}{Suggestions} can be deferred. If zero findings are produced, the operator may choose to skip Phase~3 and advance directly to Phase~4.

Beyond individual findings, the matrix supports concentration analysis --- reading the distribution of findings rather than the findings themselves. The gate requires it before advancing. Two patterns are treated as diagnostic:

\begin{itemize}[leftmargin=*]
  \item \textbf{Concentration by module.} A module drawing findings from most or all lenses is more likely conceptually flawed than locally deficient. The indicated response is redesign, not incremental patching.
  \item \textbf{Concentration by lens.} A lens drawing findings across most or all modules points to a systemic cause --- one architectural decision propagating a defect throughout the design.
\end{itemize}

\textbf{[O]} Both readings are heuristics distilled from practice, not results. They rest on an argument rather than evidence: lenses are constructed to be non-overlapping (each detects a failure class no other detects, Section~\ref{sec:lensmodel}), so if that construction holds, a module failing across independent criteria is failing for a reason those criteria have in common --- which is unlikely to be local. The argument inherits the weakness of its premise. Lens independence is asserted and untested (RO3); if two lenses overlap in practice, concentration along that pair means less than it appears to.

Neither pattern has an established threshold. ``Most or all lenses'' is deliberately vague because no principled cut exists: the diagnostic identifies where to look, and the operator judges what it means. Treating a concentration count as a decision rule would give it more authority than it has earned.

Concentration patterns are recorded as explicit decisions before Phase~3 begins, so that simplification addresses root causes rather than surface symptoms.

\section{External verification model}\label{sec:verification}

\subsection{Process architecture with gates}

Iterative Convergence institutes a sequence of phases $P_0, P_1, \ldots, P_7$, each with entry specification, expected artifact, and exit gate (Table~\ref{tab:gates}):

\begin{table}[ht]
\small
\begin{tabularx}{\textwidth}{>{\raggedright\arraybackslash}p{2.2cm} >{\raggedright\arraybackslash}p{3.2cm} >{\raggedright\arraybackslash}p{3.2cm} >{\raggedright\arraybackslash}X}
\toprule
\textbf{Phase} & \textbf{Gate} & \textbf{Primary VA} & \textbf{What it verifies} \\
\midrule
0. Discovery & G\textsubscript{0}: Score $\geq$90/100 & VA-human & Problem understanding \\
1. Architecture & G\textsubscript{1}: Design review & VA-human & All modules identified; interfaces between modules defined; no open questions that would block Phase~2 \\
2. Critique & G\textsubscript{2}: Coverage matrix & VA-human + GA (7 universal + up to 12 conditional lenses) & Critique complete; criticals recorded for Phase~3; decision for importants \\
3. Simplification & G\textsubscript{3}: Complexity audit & VA-human & Complexity $\leq$ previous version \\
4. Convergence & G\textsubscript{4}: Convergence gate & VA-human & Structural change $<$15\% vs.\ prior version [D, see §3.2]; zero critical findings; all important findings resolved or explicitly deferred with rationale \\
5. Code & G\textsubscript{5}: Compile + lint & VA-automatic & Syntax, types, style \\
6. Tests & G\textsubscript{6}: Test suite + manual & VA-automatic + human & Functional correctness + UX; spec coverage verified (a passing test is not equivalent to a verified spec criterion unless it tests the exact stated condition) \\
7. Post-Review & G\textsubscript{7}: Retrospective & VA-human & Lessons + method evolution \\
\bottomrule
\end{tabularx}
\caption{Gate structure and verification agents per phase.}\label{tab:gates}
\end{table}

\subsection{Definitions}

\begin{itemize}[leftmargin=*]
  \item \textbf{Generative Agent (GA):} The LLM. Produces textual artifacts without verification capability.
  \item \textbf{Verification Agent (VA):} Any entity capable of evaluating an artifact's correctness. Two types: \emph{VA-automatic} (unit tests, linters, compilers --- binary verdict on specific properties) and \emph{VA-human} (the operator --- evaluates semantic adequacy, usability, domain correctness).
  \item \textbf{Gate ($G_k$):} A discrete point where progression is conditioned on VA approval.
  \item \textbf{Operator:} The human practitioner directing the methodology --- the developer, architect, or domain expert who approves gates, arbitrates trade-offs, and provides domain knowledge the GA cannot verify internally.
\end{itemize}

This model should be distinguished from the \emph{dual-agent} paradigms in the 2024--2025 AI engineering literature \citep{Hovsepyan2024, Hasan2025}, which apply generator/verifier separation at the level of automated code generation and formal verification. The GA/VA model in IACDM operates at the level of \emph{process architecture}: the separation is between human-in-the-loop design phases and discrete approval gates, not between two automated LLM agents. The contribution is organisational, not computational.

Formally:
\[
G_k: \text{artifact}_k \to \{\text{approved}, \text{rejected}\}
\quad \text{where} \quad G_k = VA_1(\text{artifact}_k) \wedge VA_2(\text{artifact}_k) \wedge \ldots \wedge VA_n(\text{artifact}_k)
\]

When $G_k = \text{rejected}$, the artifact is returned to the GA with VA feedback:
\[
GA(\text{original\_prompt} + \text{VA\_feedback}) \to \text{artifact}_k'
\]

The GA still does not verify --- but now generates conditioned on concrete error information, qualitatively superior to the original prompt.

\subsection{Model properties}

\begin{enumerate}[leftmargin=*]
  \item \textbf{Generative invariance.} The GA's capability is identical in structured and unstructured processes. The methodology improves the process, not the model.
  \item \textbf{Observability.} Each gate makes observable properties of the artifact that would be invisible without external verification. The number of observable properties increases with VA diversity.
  \item \textbf{Temporal error localization.} In direct prompting, errors are detected at $t = \text{production}$. In structured process, errors are detected at $t = \text{gate}_k$, where $k$ is always prior to delivery.
  \item \textbf{Reconditioning quality.} VA rejection feedback constitutes empirical evidence of failure, conditioning next generation above the original prompt --- a property that is intuitive but not yet empirically measured (see limitation L2, Section~\ref{sec:limitations}).
  \item \textbf{Cost-tier switch point.} Gate $G_4$ marks a natural boundary for changing models. Phases~0--4 require deep reasoning over incomplete and ambiguous information; model capability is at a premium. After $G_4$, the architecture is validated and all decisions are persisted in \texttt{specs/} --- Phases~5--7 operate on well-specified, bounded tasks. This much is a consequence of the gate structure: context is externalized, so model identity matters less after $G_4$ than before it.
\end{enumerate}

\paragraph{A scope condition on Property 5, found by implementing it.}
Property~5 was originally stated with a direction: switch \emph{down} to a cheaper model
after $G_4$. Building the local-model variant showed the direction is not part of the
property. When inference runs on local hardware, tokens cost electricity rather than
per-token fees, and the economic pressure inverts --- toward spending the \emph{more}
capable model on Phases~0--4, where a local model is weakest and where an architectural
error is most expensive to discover later. The same gate structure yields the opposite
recommendation under a different cost regime.

What survives is the structural claim: $G_4$ is where externalized context makes model
identity substitutable, so it is the natural place to change models. Which direction to
change is a function of the cost regime, not of the method. The original statement
generalized from a single pricing model and was wrong to do so --- an instance of the
implicit-assumption antipattern (Section~\ref{sec:antipatterns}, AP4) in this document's own text.

\subsection{Adoption criterion}

The methodology is justified when the cost of gates is lower than the expected cost of undetected errors. Conceptually:
\[
\sum_k C_{\text{gate}_k} < \sum_j P(\text{error}_j) \times C_{\text{error}_j}
\]
This formulation is illustrative rather than directly computable: $P(\text{error}_j)$ requires historical defect data that practitioners rarely have in advance, and $C_{\text{error}_j}$ depends on detection time --- which is precisely what the methodology influences. The practical reading is qualitative: gate cost is bounded and predictable; undetected error cost is unbounded and late. See limitation L3 (Section~\ref{sec:limitations}) for the measurement agenda.

COCOMO II research suggests that early phases such as design represent a minority of total project effort but have outsized influence on final cost --- errors undetected at design phase propagate and multiply through implementation \citep{Boehm2000}. For domains where $C_{\text{error}}$ is high (healthcare, education, industrial R\&D), the inequality tends to be satisfied even with significant gate cost. The criterion is evaluable \emph{a priori}, before knowing the project size.

\section{Reference implementation: enforcement and instrumentation}\label{sec:impl}

The preceding sections describe the method. This section describes what happens when the
method is implemented as software that the generative agent cannot talk its way past ---
because that turns out to be where the method differs most sharply from its antecedents.

The distinction matters for a specific objection. External verification through gates is
not new: code review, QA sign-off, and red teaming all predate this work by decades, and
Section~\ref{sec:traditions} credits them. What is new is not the presence of a gate but its
\emph{enforceability against the agent being gated}. In human process, a gate is a social
convention: it can be waived, hurried, or declared satisfied by the party it constrains.
When the constrained party is a stochastic generator with a documented tendency to affirm
whatever it is shown \citep{Sharma2023, Perez2023}, a social gate is precisely the wrong
instrument. The implementation makes the gate a state machine external to the model.

\subsection{The gate as executable veto}\label{sec:veto}

Phase advancement is a function that can fail. A transition request is refused when any
required exit criterion for the current phase is unmarked, or when any safeguard --- one of
the eight standing checks catalogued in Section~\ref{sec:safeguards} --- applicable to that
phase evaluates to \emph{violated}. Only two transition shapes are
legal at all: advance by one phase, or the 2$\leftrightarrow$3 critique--simplification
loop. The agent may request; the state machine decides.

\subsection{Evidence gates: criteria an assertion cannot satisfy}\label{sec:evidence}

A gate that refuses transitions is only as strong as what it inspects. In the first
implementations every exit criterion was a boolean the agent set, and the state machine
checked the booleans. That is enforcement of \emph{bookkeeping}, not of work: an agent that
marks a criterion met has satisfied the gate, whatever did or did not happen.

The corpus made the gap concrete. Phase~2 requires a coverage matrix; across the recovered
projects the artifact is almost entirely absent while the criterion is marked met
throughout. Nothing was circumvented --- the criterion never required the document to exist.

Two stricter criterion classes now exist, and the distinction between them matters.

\textbf{Artifact criteria} name a file that must be present and non-trivial under
\texttt{specs/}. The architecture document, the coverage matrix, and the post-review
lessons are gated this way: the transition is refused, naming the missing path, until the
file exists. This makes the record survive the session and remain inspectable afterwards
--- which is what the corpus lacked.

\textbf{Witness criteria} are stronger, and are the more interesting construct. Phase~6's
\emph{tests passing} can no longer be asserted at all. It is satisfied only when the
process hook that observes test execution has recorded a green run \emph{in that phase};
the agent has no tool that writes this field. Where an artifact criterion asks the agent to
produce something, a witness criterion asks a separate process whether an event occurred.
This is safeguard S4 --- \emph{never assume tests passed; execute and verify} --- turned
from an instruction into a precondition.

The witness construct also exposed a defect invisible from inside either process. The MCP
server read project state once at startup, while hooks write it from separate processes, so
the server could hold a stale view and would never observe the recorded pass. Enforcement
distributed across processes requires a coherent view of state, and the fix --- reload
before each tool call, treating disk as authoritative --- is a reminder that a gate can
fail by not seeing rather than by not checking.

\paragraph{What this does not achieve.}
The engine verifies presence and non-triviality, never quality. A coverage matrix of the
right shape and the wrong content passes. Semantic adequacy remains VA-human by design
(S4), and a mechanical check that attempted to judge quality would manufacture exactly the
false assurance being corrected. Nor is any of this unforgeable in a strong sense: an agent
with filesystem access could write a hollow document or edit the state file directly. The
claim is narrower and still worth something --- the cheapest path to passing the
gate is now to do the work, where before it was to mark a boolean.

The general form of the criterion, which the validation experiment later supplied from a
failure of this very layer (Section~\ref{sec:ro3instr}):

\begin{thesis}
A gate whose integrity depends on the good faith of the party it polices is an instruction
wearing the appearance of a lock.
\end{thesis}

\paragraph{And a qualification the experiment forced.}
Hardening a gate is not sufficient, and can be actively harmful. In the validation batch the
witness gate blocked four projects whose test suites were green, because the classifier
could not read their output. The agents unblocked themselves, and disclosed doing so. Those
workarounds were not dishonesty: they were the rational response to a gate that refused with
the criterion satisfied and offered no legitimate way forward.

The lesson generalises past this implementation. \textbf{Fixing forgeability without fixing
false refusal produces stuck agents rather than honest ones} --- a lock is only a lock if
the door it guards opens when the work is done. Every enforcement mechanism here therefore
carries a second obligation beside refusing: to report what it observed, so that a refusal
can be told apart from a malfunction.

\subsection{Exit criteria}

Each phase carries explicit exit criteria --- 39 in total across the eight phases --- and
they are more demanding than this document has so far described. Phase~0, presented in
Section~\ref{sec:discoveryloop} as ``score $\geq$90 plus operator confirmation'', in fact carries ten
criteria, two of which are substantive commitments not derivable from the score at all:
\emph{technical feasibility verified} (the target platform's essential capabilities
confirmed rather than assumed --- the failure mode of porting projects) and
\emph{implementation feasibility} (every component assessed against the tier scheme of
Section~\ref{sec:tier}, with a proof-of-concept required before Phase~1 for complex tier-3
components). Phase~2 requires that every \emph{non-activated} conditional lens be recorded
with its justification, which converts lens selection from a silent omission into a
documented decision.

\subsection{Safeguards}\label{sec:safeguards}

Alongside the antipattern catalogue --- which describes failures to avoid --- the
implementation carries eight safeguards, which are checks that run. Five are validated
mechanically against project state; three constrain behaviour and are surfaced to the
agent as standing obligations rather than evaluated.

\begin{center}
\begin{tabular}{llp{7.6cm}}
\toprule
\textbf{\#} & \textbf{Enforcement} & \textbf{Constraint} \\
\midrule
S0 & mechanical & No advance past Phase~0 below the convergence score. \\
S1 & mechanical & Simplification must not introduce defects; features survive each Phase~3. \\
S2 & behavioural & \textbf{The stopping criterion belongs to the operator, not the AI.} \\
S3 & behavioural & When in doubt, iterate: stopping early is the costlier error. \\
S4 & mixed & \textbf{VA-human is irreplaceable at gates $G_0$, $G_2$, $G_4$, $G_6$.} Automated tests verify formalizable properties; semantic adequacy, usability, and domain correctness require human judgement. The \emph{execute and verify} half is now mechanical --- a witness criterion (Section~\ref{sec:evidence}) blocks Phase~6 until a passing run has been observed. The judgement half cannot be. \\
S5 & mechanical & Phases 2--3 operate strictly within Phase~0 scope. The AI may suggest; scope decisions belong to the operator. \\
S6 & mechanical & Do not reimplement what already exists (Section~\ref{sec:tier}). \\
S7 & mechanical & Sequence discipline during implementation: mark each unit complete, name the next, proceed. \\
\bottomrule
\end{tabular}
\end{center}

S2 and S4 deserve emphasis because of what the software deliberately does \emph{not}
automate in them. Both assign authority to the human operator, and automating that
authority would dissolve the property the method exists to preserve: a methodology that
decided its own stopping criterion, or that certified its own semantic adequacy, would have
reintroduced the verification gap one level up.

The boundary is not between safeguards but \emph{within} them. S4's execute-and-verify
obligation is a factual question --- did a test run pass --- and factual questions are
exactly what a witness criterion can settle without human judgement. S4's adequacy
obligation is not factual in that sense, and no mechanism here attempts it. Automating the
first strengthens the second by removing the easiest place to substitute assertion for
evidence.

\subsection{Tier assessment and blind-loop detection}\label{sec:tier}

S6 addresses a failure mode specific to generative agents: producing plausible code for a
problem that already has a verified solution. Every component is assessed into one of
three tiers --- \textbf{tier~1}, a mature library exists, so use it; \textbf{tier~2}, a
documented algorithm exists, so port it literally, preserving structure and names and
testing against the same inputs; \textbf{tier~3}, neither, in which case a bounded
proof-of-concept precedes commitment. Tier assessment is an exit criterion of Phase~0, not
advice.

The tier scheme is paired with a mechanical detector for the behaviour it is meant to
prevent. Consecutive \emph{failing} test runs with no diagnosis recorded between them
accumulate; past a small threshold, the next run is refused. The counter is cleared by two
things: a passing run, or a recorded diagnosis naming the cause. The asymmetry is
deliberate. Debugging with a named cause each round is the normal cycle and is never
blocked; failing repeatedly without naming a cause is thrashing, and is exactly the
pattern a generator will sustain indefinitely because each attempt looks locally
plausible. \textbf{[D]} The threshold is a small integer chosen to be forgiving of ordinary
debugging while still terminating a loop before it consumes a session; it is not derived
from data.

Passing runs are never blocked, because blocking one would make S4's execute-and-verify
obligation unsatisfiable --- a constraint discovered by implementation, not by design.

\subsection{The persisted process record}\label{sec:record}

Each project accumulates a machine-readable record: every phase transition with its
timestamp and the criteria satisfied at that moment; the Phase~0 score with its
per-criterion breakdown; every recorded decision with phase, category, and timestamp;
exit-criterion states; safeguard evaluations; and the undiagnosed-failure counter.

This makes several quantities recoverable that Section~\ref{sec:limitations} lists as unmeasured: wall-clock
duration per phase, the number of 2$\leftrightarrow$3 iterations actually run, the
trajectory of the convergence score, and the incidence of blind-loop episodes. The
instrument for L2 and L3 therefore already exists and has been running in production
across the projects described in Section~\ref{sec:emplimits}.

\paragraph{An earlier retrospective analysis, and why it was withdrawn.}
Twenty-seven records predating the criterion classes of Section~\ref{sec:evidence} were
recovered and analysed. That analysis is not reported here, and the reason is the same one
that motivates this section.

The tooling in force during that period had defects since corrected: exit criteria naming
an artifact did not require the artifact, so Phase~2 matrices were generally never written;
the MCP server read project state once at startup while hooks wrote it from separate
processes; and the Phase~2 lens declaration was not re-examined when the 2$\leftrightarrow$3
loop returned, so the lens set stayed fixed against V(1) while the architecture moved. The
records also mix instrument versions, since the extension changed repeatedly across the
period they span.

An argument can be constructed that phase transitions escaped these particular defects, and
it was. It is not good enough. Publishing figures whose instrument is known to have been
faulty, defended by an argument about which faults touch which code path, would trade one
transparency failure for another --- and the corpus offers no independent record to check
that argument against.

What replaces it is Section~\ref{sec:ro3}: twelve projects run under a single frozen
instrument, with the analysis plan registered before collection. The retrospective figures
are withdrawn rather than qualified, and claims this document previously supported with
them --- iteration counts, the share of time spent in design --- revert to what they were
before either was measured: recollection, unverified.

\subsection{Availability and the limits of adoption evidence}

The methodology is packaged as prompt-based protocols usable with any assistant, and as
Visual Studio Code extensions targeting different agents: Claude, Qwen, Kimi, local models
served through Ollama, and --- until it was discontinued --- GitHub Copilot. Each provides
an MCP server exposing the state machine as tools, a phase-aware sidebar, and, where the
host supports them, process hooks that observe regardless of the agent's cooperation. The
most consequential hook refuses task completion while the project fails to type-check,
lint, or pass its tests, and fails open when no toolchain is detected rather than blocking
silently.

\paragraph{What happened when the gate became advisory.}
\textbf{[O]} The Copilot build is the informative case, and it is informative because it
failed. That host provided no hook API at the time, so mechanisms the other builds place
in hooks had nowhere to run: by the author's assessment the majority of the enforcement
layer was inert, leaving the methodology dependent on the model choosing to comply. The
observed consequence was that the model self-reported compliance --- it announced
that criteria were met and phases complete, and the process had no independent means of
contradicting it. That is antipattern AP1, validation theater (Section~\ref{sec:antipatterns}),
occurring not inside a project but one level up, in the enforcement of the methodology
itself. The build was discontinued in May 2026 rather than shipped as a weaker variant.

The distinction this document draws between an enforceable gate and an advisory one is
therefore not purely analytic. Removing enforceability from an otherwise identical
implementation --- same phases, same criteria, same safeguards, same prompts --- produced
a recognisable and named failure mode, and the response was to withdraw a published
product rather than accept it. That is a costly judgement, which makes it better evidence
than a caveat would be.

Three qualifications, in keeping with Section~\ref{sec:status}. This is one case, not a
comparison: no controlled trial was run against the hook-enabled build. The proportion of
enforcement rendered inert is the author's estimate, not a measurement. And the judgement
that the advisory variant was not worth distributing was made by the method's proponent,
with all that implies --- though it is worth noting that this is a negative result about
the author's own published product, a direction in which proponent bias does not push.
Its testable form is H9.

The host capability that was missing has since been added: Visual Studio Code shipped
agent hooks in preview during 2026, and the architectural blocker no longer holds. Resuming
the build is now a porting decision rather than an impossible one, deliberately deferred
while the hook interface remains in preview.

\paragraph{The local-model build, for contrast.}
It faces the same absence of a hook system and was \emph{not} discontinued, which locates
the boundary more precisely. Enforcement there moves into the MCP layer, where the phase
transition remains a hard refusal because the state machine lives in the server rather
than in the model. What is lost is the observational layer --- a hook fires whether or not
the agent cooperates, while a tool must be called. Partial enforcement retained the gate;
the Copilot case lost it. The relevant variable is not the presence of hooks but whether
any mechanism outside the model can refuse.

\paragraph{Tool-agnosticism, at the level where it is actually demonstrated.}
The claim that the method is independent of the tool has an implementation-level reading
that is verifiable now, and an outcome-level reading that is not. At the implementation
level: the files defining the phases, exit criteria, and safeguards are \emph{byte-identical}
across the Claude, Kimi, and Qwen builds (\texttt{phases.ts}, 9{,}855 bytes, MD5
\texttt{82ABA888}; \texttt{safeguards.ts}, 7{,}543 bytes, MD5 \texttt{F9457480}), as is the
state engine between two of them. Only the adapter layer differs, and all builds share a
state format, so a project can move between agents mid-methodology. The method is one
artifact running on four model backends.

That demonstrates the method is \emph{implementable} tool-independently. It does not
demonstrate that it \emph{helps} equally regardless of model --- a different claim,
requiring the method's effect to be measured within each agent and compared across them.
That comparison has not been made, and comparing outcomes between agents without a
per-agent control would measure model capability rather than tool-independence. The
outcome-level claim is H5's neighbour and remains open.

\paragraph{On installation counts as evidence.}
Install counts for the extensions are publicly visible and indicate reach beyond the
originating institution. They should not be read as more than that: an installation is not
a use, a use is not a completed project, and none of them is evidence that the methodology
helped. Adoption figures answer ``did this leave the building'', not ``does it work''.

\section{Validation experiment: are the lenses orthogonal?}\label{sec:ro3}

The entire lens taxonomy rests on one principle: \emph{a lens is legitimate only if removing
it exposes a class of failure that no other lens detects} (Section~\ref{sec:lensmodel}).
That same section concedes what the adjacent literature found when it asked the analogous
question --- replications of Perspective-Based Reading found no significant difference
between reading perspectives \citep{Regnell2000} --- so the prior expectation here was that
the lenses overlap substantially.

This section reports an experiment designed to test that, with the analysis plan fixed
before collection.

\subsection{Corpus and procedure}

Twelve projects, each 8--12 modules and a single working session, with domains chosen so
that every conditional lens would be exercised in at least three projects --- a designed
sample, not a convenience one. All twelve ran under a single instrument version --- \texttt{versus-claude 0.14.2},
the only one installed. Every figure in this section, the lens triggers of
Section~\ref{sec:lensmodel} included, describes that version.

\begin{center}
\begin{tabular}{lr}
\toprule
Valid projects & 12 \\
Discarded, with recorded reason & 7 \\
Findings & 1{,}100 (195 critical, 657 important, 248 suggestions) \\
Distinct defects (generator clustering) & 1{,}029 \\
Modules & 130 \\
Elapsed & 37 h; 1.1--5.3 h per project, median 3.3 \\
2$\leftrightarrow$3 loop iterations & 28 total \\
\bottomrule
\end{tabular}
\end{center}

The seven discards are reported rather than dropped, as the pre-registered plan requires.
Most were discarded not for bad data but to keep the instrument constant: when a defect was
found and fixed mid-batch, projects run under the older build were withdrawn rather than
pooled. Two were discarded because an instruction without a mechanical gate failed to change
behaviour --- itself a result, reported in Section~\ref{sec:ro3instr}.

Every conditional lens cleared the three-project floor. Coverage ranged from 12 projects
(Resilience, UI/UX, Sustainability, Process, Governance, Control, Linguistics) down to 4
(Migration).

\paragraph{Availability.}
The twelve project records, their coverage matrices, the analysis pipeline, the per-project
closure reports, and the operation log recording the seven discards with their reasons are
deposited at \href{https://doi.org/10.5281/zenodo.21908908}{\texttt{doi:10.5281/zenodo.21908908}}.
That identifier resolves to the exact snapshot from which every figure in this section was
computed; \texttt{doi:10.5281/zenodo.21908907} resolves to the most recent version, which
may differ.
The elapsed figure is \texttt{createdAt} to \texttt{updatedAt} and includes operator pauses.
It is not effort, and no effort claim is made from it.

\subsection{Principal result: no lens is redundant}

\textbf{No lens among the nineteen showed zero exclusive contribution.} Every lens that
activated produced at least one defect no other lens found, in every project where it
activated.

Stated as a binary verdict this is weak, because the pre-registered removal criterion
requires \emph{all} of a lens's defects to be shared --- a bar almost nothing reaches. The
result is the distribution:

\begin{center}
\begin{tabular}{lr}
\toprule
Mean overlap across the 19 lenses & 11\% \\
Lowest (Architectural) & 2\% \\
Highest (Sustainability) & 33\% \\
Lenses above 15\% overlap & four \\
Lens pairs sharing any defect & 41 of 171 (24\%) \\
Largest pairwise Jaccard (Performance $\times$ Sustainability) & 0.10 \\
\bottomrule
\end{tabular}
\end{center}

\paragraph{A prediction of the method's own theory failed.}
The analysis plan named Assumptions $\times$ Architectural as the pair most likely
to overlap, on the intuition that asking ``what does this design assume?'' and ``can these
modules be separated?'' would surface the same defects. Measured Jaccard: 0.00.
They share nothing. A prediction that could have failed, made by the people who wrote the
method, failed against them --- which carries more weight than the confirmations.

\paragraph{Robustness: four independent clusterings.}
Exclusive contribution depends on who decides that two findings are the same defect. That
judgement was redone blind by three independent evaluators on a package that
removes the lens column, renames finding identifiers (whose prefixes leak the lens),
shuffles the order deterministically, and deletes the original duplicate marks.

Under the most aggressive reading --- the \emph{union}, in which any pair that
\emph{any} evaluator merged counts as one defect --- 1{,}029 defects collapse to 668, a 35\%
reduction. \textbf{No lens reaches zero under any of the four clusterings.} The lowest is
Migration at 3, and it activated in only 4 of 12 projects.

The effect is not monotonic: four lenses \emph{gain} exclusive contribution under one
clustering, because when other clusters merge, a lens can become the sole occupant of
ground it previously shared.

\paragraph{What this does not establish.}
Not that nineteen is the right number, that the set is complete, or that a different
division would be worse. Only that none is redundant in this corpus under the
registered criterion. Completeness was not tested.

\subsection{Secondary result: the activation criterion is the weak part}

This was not in the design. It emerged from measuring whether lens activation is
reproducible, and it is the more consequential of the two.

Each conditional lens carries a \emph{central question} and a \emph{trigger}. In at least
five lenses these describe different kinds of thing, and different readers answer different
ones. Phase~2 applies the central question; independent readers apply the trigger.

The evidence is textual rather than statistical --- blind estimators quote the trigger
verbatim when declining a lens:

\begin{itemize}[leftmargin=*]
  \item \textbf{Ethical:} trigger reads ``automated decisions \emph{about people}''; the estimator declined with \emph{``classifies X.509 certificates, not people''} --- correct against the trigger, wrong against a central question about human impact.
  \item \textbf{Linguistics:} trigger reads ``interface contracts \emph{between independent teams}''; declined with \emph{``internal contracts in TypeScript in the same repository''} --- while the central question, about contract ambiguity, plainly applied.
\end{itemize}

\paragraph{The finding survived a reader that agrees 91\% of the time.}
Four estimators re-derived the lens set blind from architecture V(1), three rounds each.
Agreement with the Phase~2 declaration ranged from 74\% to 91\%. Divergences do not scatter:
they concentrate in six lenses, and 35 of them run in a single direction --- for
Sustainability, Resilience, Control, Ethical and Linguistics the external readers
consistently decline what Phase~2 activated, while Mechanical is the mirror. Systematic
directional disagreement is not noise; it is two readings of one criterion.

\paragraph{A natural explanation was tested and refuted.}
One estimator ran on a locally-hosted quantised model, inviting the conclusion that
quantisation explained its weaker agreement. The same model at full precision scored
\emph{lower} (74\% against 81\%), with equivalent instability, and the two agree with each
other 92\% of the time --- they are the same reader. The difference between estimators is
model family, not numerical precision. This removes an easy explanation that would have
weakened the result.

\paragraph{A second mechanism: late emergence.}
In 3 of 12 projects the lens set changed \emph{between} loop iterations, always with a lens
entering against a later architecture. This is not a property of any particular lens --- an
early hypothesis that one lens was structurally late was refuted --- but of how much a given
architecture moves under critique. It matters for coverage counting: how many lenses a
project exercises depends partly on how many loop turns it took, which is independent of its
domain.

Re-declaration is real work rather than ritual: across all twelve projects, non-activation
justifications were compared between iterations and no two were identical.

\subsection{Result on the method: commitments are never re-confronted}\label{sec:ro3method}

The third result concerns the method rather than the taxonomy, and emerged from the
implementation phases.

The method creates two kinds of binding commitment --- \emph{premises} recorded in Phase~1,
and \emph{resolutions} of critical findings recorded in Phase~3 --- and \textbf{nothing ever
checks that the built system honours either}. Four occurrences across the batch, by four
independent mechanisms: an eradicated complexity class reintroduced in three places; an
approved refinement never implemented; a premise of streaming iteration contradicted by
memory growth linear in input; a single-thread premise contradicted by a framework running
handlers in a thread pool.

All four surfaced by favourable accident --- a timer in an unrelated test, a smoke
test that happened to exist, a test of a different property, a database error. None was
found by a mechanism designed to find it.

The batch produced its own criterion for the fix:

\begin{thesis}
A premise is protected only when a test exists that would measure it failing.
\end{thesis}

The correction is cheap because the linkage already exists: premises are numbered,
findings carry identifiers, resolutions are recorded. Only the check is missing.

\paragraph{Five ways a passing test fails to test.}
A sibling catalogue emerged: the test exercises a scenario that does not reach the criterion;
the fix closes one path while the defect returns by another; the condition the test would
exercise cannot occur; the invariant holds whether the premise is true or false; and the
test fails for reasons unrelated to any product defect.

The sharpest is the third. A guard was implemented, tested, and green --- and with the window
size equal to the minimum sample and a bounded queue of the same length, the condition it
checked could never become false. Sixty-two passing tests did not notice; mutation
testing did, and the symptom was that disabling the check broke only 2 tests where another
mutation broke 14.

No safeguard in the methodology covers this class. The countermeasures that worked
were invented by the agents, unprompted: measuring per stage before optimising, verifying by
execution rather than reading, mutation testing, and deriving a test map from the
specifications. Each appeared in three or four independent projects. When independent runs
converge on the same unprescribed practice, the practice belongs in the method.

\subsection{What the experiment revealed about its own instrument}\label{sec:ro3instr}

\paragraph{The witness gate is forgeable, and failed in a third of projects.}
The hook recording test outcomes failed to register in 4 of 12 projects. In all four the
agent unblocked itself by feeding the classifier real output from a real run, and
in all four disclosed the workaround unprompted. No falsification occurred. The
defect is that the mechanism would not have prevented one: the hook accepts any JSON on
standard input, and the recorded state carries no provenance.

Three distinct causes were identified, and the third defeats the simple explanation: output
markers not matching one reporter's word order; markers that \emph{did} match while the hook
was never invoked; and a duplicated quiet flag suppressing the summary, making a green suite
unclassifiable. Two design defects are demonstrated --- classifying prose instead of exit
codes, and \emph{not observed} being indistinguishable from \emph{observed and failed}.
Why the hook did not fire remains undetermined, and is stated that way deliberately.

\paragraph{Exit criteria record the assertion, not the state of the mechanism.}
Exit criteria were substantial: 390 records, 791 characters on average, none empty. Yet in
the four projects where the gate was bypassed, the tests-passing criterion recorded the
result correctly and never mentioned that the mechanism had failed. An auditor
reading only the criteria would conclude the gate worked in all twelve.

\begin{thesis}
Every method defect in this experiment came from the narrative log, not from the exit
criteria. The criteria record compliance accurately; the narrative records what went wrong
on the way.
\end{thesis}

\paragraph{An instruction without a gate does not change behaviour.}
Four occurrences, two of which cost a discarded project. This is empirical validation of
evidence-gating (Section~\ref{sec:evidence}) from the inside. The instructive case is the
one that \emph{cannot} be gated: recognising that two findings describe the same defect is a
quality judgement, mechanically undecidable --- and the inter-rater figures below are the
empirical proof. It marks the boundary of the principle rather than an application of it.

\subsection{What the batch produced besides results}\label{sec:ro3patch}

An experiment that runs a method twelve times under observation yields more than the
measurement it was designed for. This one produced a corrective specification for the
instrument: fourteen changes across the enforcement layer, each naming the exact
code surface, the corpus evidence that motivated it, and its cost.

That specification is governed by a rule fixed before the batch and applied to every item:
a guidance instruction becomes a mechanical gate only when it is a property of \emph{form}
--- existence, membership in a closed set, structural reference --- \emph{and} has an
observed, correctly diagnosed failure in the corpus. Quality properties are excluded by
construction, because they are the oracle problem and no mechanism decides them.

Six of the fourteen address the witness gate alone, and they are specified as a package
for the reason given in Section~\ref{sec:evidence}: correcting forgeability without
correcting false refusal would produce blocked agents rather than honest ones.

One item illustrates the class of defect that only sustained operation surfaces. When the
outcome classifier could not read a test run it exited without recording anything, while the
gate read recorded state rather than execution. The composition fails closed and
\emph{permanently}: once a failure is recorded, any later unreadable run leaves it standing,
and re-running the suite green changes nothing. No single reading of either component shows
this; it appears only when both run against a real project.

This is what the phase was for. A method that produces a specified correction of
its own enforcement layer --- with code surfaces, evidence, and costs --- has been tested
differently from one that produces an opinion about itself. The specification is also the
answer to why the operator was the method's author for this batch, taken up in
Section~\ref{sec:strategy}: recognising a defect on sight is a precondition for finding
these, and it is incompatible with blinding.
\subsection{Measured limitations}\label{sec:ro3limits}

These were measured, not declared, and the section is not shortened.

\paragraph{Duplicate marking is imprecise, not irreproducible.}
This corrects a reading circulated mid-batch. Across 51{,}565 evaluable pairs, pairwise
Cohen's $\kappa$ between the four evaluators ranges from 0.110 to 0.338 --- every
pair \emph{dozens of times above chance}. What varies is the marking \emph{rate}, not the
judgement: the generator marked 71 pairs, the strongest external reader 340. That reader
recovered 63\% of the generator's marks reading blinded descriptions alone, and 27
pairs were confirmed by three evaluators independently.

Two consequences. The hypothesis that the generator perceives relations unrecoverable from
the text is \textbf{eliminated} --- a blind reader recovers two thirds of them. And the
hypothesis that ``same defect'' is not definable at this granularity is \textbf{weakened}.
Reporting $\kappa$ without the marking rate beside it produces the opposite conclusion; that
error was made during this batch and corrected.

\paragraph{The quantised local model is an outlier.}
It marked 87 pairs --- more than the generator --- while sharing only 17 with its own
full-precision version. It does not mark less; \emph{it marks something else}. This is a
datum about locally-hosted quantised models in fine semantic judgement, and a methodological
warning.

\paragraph{Self-evaluation.} Findings and primary duplicate marks come from the same
generating model. Blind re-marking measures the size of that problem; it does not remove it.

\paragraph{Human verification varied in provenance.} The operator executed and judged in
eight projects, judged agent-executed results in three, and accepted without
executing in one. That project became an unplanned natural experiment: it was the only one
without human verification, and four defects surfaced after the operator's approval,
including the refuted streaming premise. Approval is not verification, and the evidence
comes from the project where verification was declined.

\paragraph{Disclosed harness workarounds.} Four of twelve, all self-disclosed. The frequency
is data; the integrity of the record depended on good faith rather than on the mechanism.

\paragraph{Scope.} One generating model, twelve projects of comparable size, domains selected
to exercise the conditional lenses. Not a random sample of real software.

\subsection{Corrections made during the batch}

Seven intermediate readings were contradicted by later projects; the entry on estimator
agreement was rewritten four times. This is reported because the trajectory is part of the
result.

The design called for incremental analysis, project by project, which gave each new project
the power to refute the standing reading. Five refutations occurred --- including the claim
that duplicate marking was irreproducible, the claim that one lens activated by
architectural maturation, and the claim that quantisation explained an estimator's weakness.
A result that survived that is more trustworthy than one computed once at the end.

The expectation registered before the third evaluator ran was contradicted by the
data. The stated prediction was that the construct would prove undefinable; the measurement
weakened that prediction rather than supporting it. Registering an expectation and then
publishing its refutation is what separates pre-registration from decoration.
\section{Antipatterns to avoid}\label{sec:antipatterns}

\begin{enumerate}[leftmargin=*]
  \item \textbf{Validation theater.} Asking the AI ``is this design good?'' and accepting affirmation. Antidote: request refutation, never validation.
  \item \textbf{Complexity inflation.} When critique reveals fragility, adding components. IACDM treats simplification as a first-class design obligation: each iteration must maintain or reduce architectural complexity relative to the previous version. Growth in component count without corresponding growth in understood requirements is a signal of design failure, not progress.
  \item \textbf{Monolithic session.} Design, code, and tests in a single conversation. Context saturates. One session per phase, one phase per module.
  \item \textbf{Implicit assumption.} Design that works only if unlisted conditions are true \citep{Leveson2011}.
  \item \textbf{VA-human absence.} Relying exclusively on automated tests. Semantic adequacy and domain correctness require human judgment.
  \item \textbf{Phase 0 bypass.} Starting with architecture without understanding the problem. Cost multiplier: orders of magnitude relative to early discovery \citep{Boehm1986}.
  \item \textbf{Reference-free implementation.} Implementing domain algorithms without bibliographic reference. AI generates plausible code with invented parameters.
  \item \textbf{Convergence perfectionism.} Infinite critique cycles seeking ``perfect'' design. The stopping criterion is defined at gate G\textsubscript{4} (Section~\ref{sec:verification}): structural change below 15\% relative to the prior version, zero critical findings, and all important findings resolved or explicitly deferred with rationale. When these conditions are met, the design is ready to evolve --- not finished, but ready.
  \item \textbf{Silent scope creep.} Adding features during Phases 2--3 without operator approval. Scope decisions belong to the human. AI can suggest; never decide.
  \item \textbf{Verified-solution bypass.} AI generates plausible code for problems with established solutions, replacing tested implementations with stochastic approximations. Antidote: tier assessment per module in Phase 1 --- if during Phases 5--6 the AI is doing trial-and-error on something deterministic, stop and search for a reference implementation.
  \item \textbf{Implementation drift.} Each module is coded in a separate session, creating risk of silent divergence between implementation and the \texttt{specs/} established in Phases~0--4. IACDM addresses this with an \textbf{adversarial micro-check} after each module: the GA is explicitly asked ``where does this implementation diverge from \texttt{specs/}?'' --- not ``is this correct?''. The adversarial framing is deliberate: the GA that wrote the code has an agreement bias toward confirming correctness (AP1); the micro-check forces it into the refutation posture that reduces this bias.
  \item \textbf{Scope presence blindness.} Passing automated tests (G\textsubscript{5}) does not verify that all required modules were implemented. Before advancing from Phase~5 to Phase~6, a \textbf{scope inventory} is required: every module defined in Phase~1 decomposition must be present in the codebase, and every requirement in \texttt{specs/validation} must be covered by at least one module. This is a presence/absence check only --- quality is Phase~6's responsibility. Modules that pass all tests but were never implemented do not fail tests; they simply do not exist.
\end{enumerate}

\section{Limitations and future work}\label{sec:limitations}

\subsection{Validation strategy: what was tested first, and why}\label{sec:strategy}

The limitations below are easier to misread as a list of gaps than as what they are --- the
boundary of a phase that was entered deliberately. Two choices govern the shape of the
evidence in this document, and both were made in advance.

\paragraph{Mechanism before outcome.}
The obvious experiment is whether IACDM outperforms direct prompting. It was not run first,
and running it first would have been the wrong order. The method's most distinctive
component is the lens taxonomy, and its legitimacy rested on an untested design principle
that the nearest empirical literature contradicted \citep{Regnell2000}. Had that principle
failed, the critique phase would not survive in its current form and an efficacy study of a
method about to be restructured would have measured the wrong thing. Section~\ref{sec:ro3}
therefore tested the mechanism. The outcome question remains open, and is H7.

\paragraph{Instrument shakedown before operator independence.}
The validation batch was simultaneously a measurement and a debugging run of the tooling
that enforces the method. It discarded seven projects to keep the instrument constant,
diagnosed a gate failure occurring in a third of runs, rewrote three lens triggers that were
producing no data, and issued the fourteen-item corrective specification of
Section~\ref{sec:ro3patch}.

None of that is available to an operator following a script. Recognising a defect on sight
--- distinguishing a genuine refusal from a malfunction, a legitimate variation from a
violation --- requires knowing what the mechanism was supposed to do. For this phase
the operator had to be the method's author, and that role is incompatible with blinding.
It was accepted knowingly, and it bounds what the phase can show.

\paragraph{What follows from having chosen this order.}
This document can report that the mechanism withstands testing, that the criterion deciding
when to apply it does not, and that the method leaves commitments unverified. It cannot
report that the method works, or that its effects survive an operator who did not design it.
Those are the next phase, and the conditions for it now exist that did not before: the
instrument is frozen, the analysis plan is written, and the institutional approval for
independent operators is in place.

The limitations that follow should be read against that: they describe the edge of a phase,
not oversights within one.
\subsection{Empirical limitations}\label{sec:emplimits}

\paragraph{L1 --- Single-institution validation.}
The corpus has two tiers, and the split is deliberate rather than incidental.

\textbf{Innovation projects} at INDT --- an industrial R\&D institute in Manaus, Brazil --- are the work the method exists to serve. They are also poor instruments for evaluating it: an innovation project is unrepeatable by construction --- if its course were predictable it would not be innovation --- it cannot be afforded twice, and no control arm can be run on it without doubling a cost the organisation is carrying for other reasons.

\textbf{An evaluation set} was therefore built alongside it: projects chosen for breadth of type and short cycle time, run specifically to exercise the method across a spectrum the institutional work does not span. Of the 23 recovered records that contain phase transitions, 3 are institutional projects and 20 belong to the evaluation set --- a fourth institutional project was abandoned at initialisation and appears only in the 27-record total. The evaluation set covers arcade-game replications, board and logic games, CRUD and small business applications, a scientific computation tool, a data-driven recommendation application, and tooling analysis. Durations range from under an hour to more than two months.

This is a stated trade-off rather than a convenience. The evaluation set buys breadth, repeatability, and cycles short enough to observe a whole methodology pass end to end; it pays in external validity, because a one-hour game replication is not a two-month innovation project and no argument offered here bridges that gap.

The tooling persists a machine-readable record per project (Section~\ref{sec:record}). Those records show the method was applied --- phases entered, gates passed, decisions recorded --- and they document \emph{adherence}, not \emph{effectiveness}: no project in either tier has a non-IACDM counterpart.

INDT has authorised release of the derived process metrics for the institutional tier, so the corpus can be reported in full rather than only for the evaluation set. Project \emph{contents} --- recorded decision text, specifications, source --- remain out of scope, since they carry client and business detail that the process metrics do not. No release has been made: the tooling defects described in Section~\ref{sec:record} were found after the analysis, and publishing figures whose instrument is known to have been faulty would trade one transparency failure for another. What is published instead is the account of what the corpus does and does not support, and the design of the batch that replaces it.

\textbf{This establishes applicability across operators and project types within one organisational context, and nothing more.} It does not establish superiority over any alternative, because no alternative was run. Institutional confounds --- shared practices, shared tooling, shared culture --- cannot be ruled out.

A third concentration compounds the first two: every project in the validation batch
of Section~\ref{sec:ro3} ran on a single generating model, as did the great majority of the
earlier work. The corpus is therefore single-institution, largely single-operator, and
single-model. It cannot distinguish ``IACDM works'' from ``IACDM with this model works''.

This does not touch the implementation-level tool-independence of Section~\ref{sec:impl},
which rests on the artifacts being identical rather than on how often each was run; it bears
on the outcome-level claim, which that section already declines to make. It does bound the
orthogonality result: the lenses are non-redundant \emph{as exercised by one model}, and
whether a different model would partition the failure space the same way is untested. The
blind estimators offer a partial hint --- four models from three families agreed 74--91\% on
which lenses apply --- but agreement on activation is not evidence about what each lens
finds once applied.

\paragraph{L6 --- The evaluators are the proponents.}
The method, the tooling that enforces it, the projects it was applied to, and this assessment of it all originate with the same author, inside one institution. Every [O] claim in this document is therefore subject to confirmation bias in a strong form: the observer chose what to notice, when to notice it, and how to describe it, while holding a stake in the outcome. Machine-recorded process data (Section~\ref{sec:record}) removes self-report from the timing and adherence figures, but it does not remove the biases that operate before recording --- which projects were attempted, which were abandoned, how difficulty was judged, when a design was declared converged.

No amount of instrumentation fixes this; only independent operators and blinded outcome assessment do.

Two qualifications, and neither dissolves the limitation. The validation experiment
(Section~\ref{sec:ro3}) moved one judgement out of the author's hands entirely: the
clustering of findings into distinct defects was redone blind by three independent
evaluators, and the principal result holds under all four. And for that batch the authorship
of the operation was a requirement rather than a concession (Section~\ref{sec:strategy}) ---
the method findings, the gate diagnosis, and the fourteen-item corrective specification came
from someone able to recognise a defect on sight.

What remains is precisely what those two do not reach: no one independent has \emph{operated}
the method, and no outcome has been assessed blind to condition. Note also what \emph{cannot} be blinded: an operator following an eight-phase gated process necessarily knows which condition they are in. The achievable design is therefore independent operators plus outcome assessment blind to condition, plus pre-registered analysis --- and until that exists, the comparative claims in this document remain hypotheses rather than results.
\emph{Testable hypothesis (H8):} the effect attributed to IACDM is smaller when measured by operators independent of the method's author than when measured by the author.
\emph{Testable hypothesis (H5):} Operators with different experience levels applying IACDM on distinct projects in independent organisations produce results of measurably superior quality --- as assessed by defect rate, rework time, or operator-reported confidence --- compared to direct prompting.

\paragraph{L2 --- Property 4 not empirically tested.}
The claim that VA error feedback produces ``qualitatively superior conditioning'' is intuitive but unmeasured.
\emph{Testable hypothesis (H6):} Defect rate decreases monotonically with VA feedback cycles at each gate, up to a stabilization point.

\paragraph{L3 --- Gate cost partially measured; overhead not measured at all.}
Per-phase durations have now been extracted (Section~\ref{sec:record}): design occupies a median 44.6\% of elapsed time in single-session projects and behaves as a near-fixed entry cost rather than a proportional tax. That refutes the 15--25\% figure this document previously carried, and it decomposes where the method spends its time.

It does not measure overhead. Overhead is the difference against \emph{not} using the method, and no project in the corpus was run that way. Two further gaps separate these figures from H7: the records capture calendar rather than active time, since the tooling emits no activity signal; and nothing is recorded after delivery, which is precisely where the compensating benefit is claimed to lie. A recording-only mode for a control arm, an activity signal, and a post-delivery defect log are prerequisites, not refinements.
\emph{Testable hypothesis (H7):} Total wall-clock time under IACDM exceeds direct prompting on matched tasks, and the difference is offset by reduced post-delivery rework measured on an acceptance suite fixed before either condition runs.

\subsection{Partially addressed limitations}

\paragraph{L4 --- VA-human competence.}
The model assumes a competent operator, but the METR study demonstrates even experienced developers confuse fluency with correctness. Binary checklists, quantitative scores, and adversarial posture mitigate but do not eliminate this bias.

\paragraph{L5 --- Team scalability.}
The method was validated with an individual operator. In teams, VA-human coordination, adversarial posture transfer, and systemic vision maintenance present incompletely addressed challenges.

\subsection{Structural limitations}

The GA/VA model does not model failure modes of the method itself when operated correctly. Additionally, the equation $E = I_0/C$ captures AI context efficiency but not the human operator's cognitive cost. The optimal granularization point --- balancing $E_{\text{AI}}$ and $E_{\text{human}}$ --- is not formalized.

\subsection{Validation roadmap}

\begin{center}
\begin{tabular}{clp{4cm}p{4cm}}
\toprule
\textbf{\#} & \textbf{Limitation} & \textbf{Validation method} & \textbf{Success metric} \\
\midrule
L1 & Single-institution (INDT); two-tier corpus & Independent replication, distinct organisations & Quality $>$ direct prompt in $\geq$2 independent organisations \\
L2 & Property 4 & Gate instrumentation & Monotonic defect convergence \\
L3 & Gate cost & Paired timing, both conditions & Overhead bounded; rework reduced \\
L4 & VA-human competence & Novices vs.\ experienced & Checklists reduce perception--reality gap \\
L5 & Team scalability & Project with team $\geq$3 & Coordination via specs/ functional \\
L6 & Proponents are the evaluators & Independent operators; blinded outcome assessment & Effect survives with operators independent of the author \\
\bottomrule
\end{tabular}
\end{center}

\subsection{Catalogue of testable hypotheses}\label{sec:hypotheses}

Every [D] and [O] claim that could in principle be wrong is restated below in refutable
form. The list is the honest inventory of what this document asserts without having
established.

\begin{center}
\begin{tabular}{lp{11.2cm}}
\toprule
\textbf{\#} & \textbf{Hypothesis} \\
\midrule
H1 & Holding the model fixed, variation in pre-generation design quality accounts for more variance in delivered software quality than variation in model capability does (Thesis~1). \\
H2 & Decomposing work into units that fit a single focused session yields fewer defects per unit of output than equivalent work issued as monolithic requests (Thesis~2). \\
H3 & Defects originating in problem misunderstanding cost substantially more to correct than defects originating in implementation, in AI-assisted settings specifically (Thesis~3). \\
H4 & For projects that turn out simple, the abbreviated Phase~0 costs on the order of tens of minutes, and that cost is small relative to the rework it averts. Neither half is measured. \\
H5 & Operators of varying experience applying IACDM in independent organisations produce measurably higher quality than direct prompting. \\
H6 & Defect rate decreases monotonically with VA feedback cycles at each gate, up to a stabilization point. \\
H7 & IACDM increases wall-clock time on matched tasks, and the increase is offset by reduced post-delivery rework on an acceptance suite fixed before either condition runs. \\
H8 & The effect attributed to IACDM is smaller when measured by operators independent of the method's author than when measured by the author. \\
H9 & Holding phases, criteria, safeguards, and prompts constant, an advisory gate (one the agent may decline to honour) yields measurably higher rates of unwarranted self-reported compliance than an enforced gate (one an external mechanism can refuse). \\
\bottomrule
\end{tabular}
\end{center}

H7 and H8 are the two that decide whether the method's central claim survives. Neither has
been tested. H8 in particular is stated in the direction that would embarrass the method,
because that is the direction a proponent is least likely to look.

\begin{thesis}
Making limitations explicit is Property 2 (observability) applied to the method itself. What is not observable cannot be corrected.
\end{thesis}

\subsection{Research opportunities}

Beyond validating what this paper claims, the framework opens a set of research questions that were not previously formulable in this form. These are not gaps to be filled before the methodology is usable --- it is usable now --- but directions that the work makes visible for the first time.

\paragraph{RO1 --- Teach-back as an independent mechanism.}
The teach-back inversion (Section~\ref{sec:discoveryloop}) is the most empirically tractable contribution of this paper. A controlled experiment with two groups --- one using direct confirmation (``did you understand?''), the other using structured teach-back --- measuring the rate of misunderstanding detected before Phase~1 begins would produce a publishable result independently of IACDM as a system. The hypothesis is sharp: teach-back surfaces a measurably higher proportion of understanding errors than direct confirmation, because it shifts the evaluation burden from the model (which will always affirm) to the human operator (who can detect deviations from intent).

\paragraph{RO2 --- The verification gap as a measurable construct.}
This paper defines the verification gap conceptually and supports it with indirect evidence (METR, Palmer, GitClear). A direct measurement is missing: the correlation between the presence or absence of structured external verification gates and defect rate in production, controlling for developer experience. The METR study comes close but was not designed for this question. A study instrumenting projects with and without gate-structured processes --- measuring defect rate, rework time, and perception--reality gap --- would give the verification gap empirical standing as a construct, not just a conceptual claim.

\paragraph{RO3 --- Lens validity: answered, and what remains.}
This was the open question of earlier versions, and it has been tested
(Section~\ref{sec:ro3}). The prior expectation, taken from the negative replications of
Perspective-Based Reading \citep{Regnell2000}, was that the lenses would overlap
substantially. They did not: no lens showed zero exclusive contribution, under four
independent clusterings, with mean overlap of 11\%.

What remains open is narrower and worth stating precisely. Completeness was not
tested --- the experiment shows no lens is redundant, not that the set is sufficient, and a
study asking which findings fall outside all nineteen has not been run. Nor was
generality across models --- one generating model produced every finding, so whether a
different model partitions the failure space the same way is unknown.

And the divergence from the PBR result needs explaining rather than celebrating. Three
candidates: perspectives assigned to one generator under explicit criteria may separate
better than perspectives assigned to different humans; architecture may partition into
failure classes more cleanly than requirements text; or the granularity at which ``same
defect'' is judged may differ between the two settings enough to move the measurement.
Section~\ref{sec:ro3limits} shows that granularity is imprecise here --- and the PBR
literature, so far as we can tell, never measured its own.
\paragraph{RO4 --- The optimal granularization point.}
The context efficiency metaphor $E = I_0/C$ (Section~\ref{sec:granular}) points to a real optimization problem: what module size maximizes output quality without exceeding the human operator's cognitive integration cost? This is a two-objective problem --- AI context efficiency and human working memory load --- and the optimal point is likely domain-dependent. Empirical measurement across project types and operator experience levels would produce actionable guidance for practitioners and theoretical constraints for the equation's formalization.

\paragraph{RO5 --- Methodology robustness under model capability improvement.}
The central argument of this paper is that the verification gap is structural and model-independent. This argument is currently unfalsified but also untested against the strongest available counterevidence: models with native tool use and self-correction capabilities. If future models can reliably verify their own output for well-specified problem classes, the verification gap partially closes --- and the scope conditions of IACDM become a research question in their own right. What problem classes remain irreducibly dependent on external verification even as model capabilities improve? This is the long-horizon question that the SE~3.0 trajectory \citep{Hassan2024} and IACDM together make urgent.

\section{Synthesis}

IACDM is a development discipline that repositions AI from code generator to design partner. It operates in 8 phases with discrete gates, sustained by three pillars: deep problem discovery before any technical solution (Phase 0, with Hierarchical Semantic Analysis), persistent knowledge management across sessions and versions (\texttt{specs/}), and systematic external verification at each phase transition (GA/VA model).

The methodology ships as Visual Studio Code extensions for four agents, with installations beyond the originating institution (Section~\ref{sec:impl}) --- evidence of reach, not of benefit. The external verification model formalizes the methodology's actual role: it does not alter the model's capability (Property 1 --- Generative Invariance). What it does is institute a framework of external verification agents --- automatic and human --- operating at discrete gates, making errors observable and providing concrete feedback for reconditioning.

What changes is not what the model can do, but what the process can see.

\paragraph{The scope of that claim, stated precisely.}
``Errors invisible to the generative agent become observable at gate boundaries'' is
\emph{analytic}, not empirical --- and saying so is the honest form of the claim. It
follows from the definitions in Section~\ref{sec:verification} and nothing else. The GA is defined as having
no internal $V(\text{artifact})$; a VA is defined as an entity that evaluates artifacts;
a gate is defined as a point where a VA does so. Given those definitions, an error that
a VA detects at $G_k$ was by construction not detectable by the GA alone. The statement
is true the way ``an external check checks things the internal one cannot'' is true. It
is a consequence of how the terms are set up, and no observation could refute it.

What that leaves open is everything that matters practically, and none of it is settled
here: whether real VAs detect a useful fraction of real errors; whether the errors they
catch are the expensive ones; whether the gates cost less than the errors they prevent;
and whether operators sustain the discipline under delivery pressure. Those are empirical,
and they are H5--H8. The analytic claim guarantees only that the architecture creates a
place where verification \emph{can} occur --- not that anything worthwhile occurs there.

Similarly, assumptions that would remain implicit become explicit artifacts subject to
critique --- but only those the operator or a lens actually surfaces. The method makes
declaration possible and demands it; it cannot guarantee completeness, and AP4
(implicit assumption) remains in the catalogue precisely because the mechanism leaks.

The methodology's contribution is therefore architectural: it converts an opaque
generation process into a cycle with designated points where verification, if performed,
is external to the generative agent. \textbf{[L]} The 2025 evidence surveyed in Section~\ref{sec:motivation}
establishes that the underlying failure is not tool-specific, which is why an answer
pitched at the process level is the appropriate kind of answer. It does not establish that
this particular process is an effective one --- that remains the open question of
Section~\ref{sec:limitations}.

\subsection{Positioning within the SE 3.0 literature}

Recent work on the next era of software engineering converges on a diagnosis: the SE 2.0 model --- AI as a task-driven copilot completing isolated code fragments --- exposes structural limitations including cognitive overload, loss of architectural coherence, and verification gaps \citep{Hassan2024}. Hassan et al. propose SE 3.0 as an intent-first, conversation-oriented paradigm in which AI evolves into an intelligent collaborator capable of understanding software engineering principles. IACDM shares the diagnosis and the directional shift, but differs in emphasis: where SE 3.0 frames the evolution as primarily a capability problem (building AI systems that reason more deeply about intent), IACDM treats it as primarily a process problem (structuring human-AI interaction so that errors become observable regardless of model capability). The two positions are complementary rather than competing. SE 3.0's vision requires that, as model capabilities improve, humans retain meaningful verification authority; IACDM operationalizes that authority through gates, lenses, and external verification agents that function independently of the underlying model. An empirical study of AI-assisted development practice with 40 software practitioners \citep{Amasanti2025} independently confirmed that asking AI assistance on small, focused problems maximizes productivity and that rework increases sharply with task complexity --- consistent with IACDM's granularization and phase-scoping principles, though derived from different methodological premises.

\begin{center}
\emph{``A converged design is not a finished design,}\\
\emph{but one that is certainly ready to evolve.''}
\end{center}

\bibliographystyle{plainnat}

\begin{thebibliography}{99}

\bibitem[Basili et al.(1996)]{Basili1996}
Basili, V.~R., Green, S., Laitenberger, O., Lanubile, F., Shull, F., S{\o}rumg{\aa}rd, S., \& Zelkowitz, M.~V. (1996). The empirical investigation of Perspective-Based Reading. \emph{Empirical Software Engineering}, 1(2), 133--164.

\bibitem[Regnell et al.(2000)]{Regnell2000}
Regnell, B., Runeson, P., \& Thelin, T. (2000). Are the perspectives really different? Further experimentation on scenario-based reading of requirements. \emph{Empirical Software Engineering}, 5(4), 331--356.

\bibitem[Hassan et al.(2024)]{Hassan2024}
Hassan, A.~E., Oliva, G.~A., Lin, D., Chen, B., \& Jiang, Z.~M. (2024). Towards AI-native software engineering (SE 3.0): A vision and a challenge roadmap. \emph{arXiv:2410.06107} [cs.SE].

\bibitem[Amasanti \& Jahić(2025)]{Amasanti2025}
Amasanti, G., \& Jahić, J. (2025). The impact of AI-generated solutions on software architecture and productivity: Results from a survey study. In \emph{Proceedings of the International Workshop on AI-Assisted Software Architecting (AISA 2025), co-located with ECSA 2025}, Limassol, Cyprus. \emph{arXiv:2506.17833} [cs.SE].

\bibitem[Argyris(1977)]{Argyris1977}
Argyris, C. (1977). Double loop learning in organizations. \emph{Harvard Business Review}, 55(5), 115--125.

\bibitem[Nuseibeh \& Easterbrook(2000)]{Nuseibeh2000}
Nuseibeh, B., \& Easterbrook, S. (2000). Requirements engineering: a roadmap. In \emph{Proceedings of the Conference on the Future of Software Engineering} (ICSE 2000), pp.~35--46. ACM Press. \url{https://doi.org/10.1145/336512.336523}

\bibitem[Beck(1999)]{Beck1999}
Beck, K. (1999). \emph{Extreme Programming Explained}. Addison-Wesley.

\bibitem[Beck(2003)]{Beck2003}
Beck, K. (2003). \emph{Test-Driven Development: By Example}. Addison-Wesley.

\bibitem[Boehm(1986)]{Boehm1986}
Boehm, B. (1986). A spiral model of software development and enhancement. \emph{ACM SIGSOFT Software Engineering Notes}, 11(4), 14--24.

\bibitem[Boehm et al.(2000)]{Boehm2000}
Boehm, B., Abts, C., Brown, A., Chulani, S., Clark, B., Horowitz, E., Madachy, R., Reifer, D., \& Steece, B. (2000). \emph{Software Cost Estimation with COCOMO II}. Prentice Hall.

\bibitem[Brooks(1987)]{Brooks1987}
Brooks, F. (1987). No silver bullet: Essence and accidents of software engineering. \emph{Computer}, 20(4), 10--19.

\bibitem[Constantine \& Yourdon(1979)]{Constantine1979}
Constantine, L., \& Yourdon, E. (1979). \emph{Structured Design}. Prentice Hall.

\bibitem[Dijkstra(1974)]{Dijkstra1974}
Dijkstra, E. (1974). On the role of scientific thought. \emph{EWD447}.

\bibitem[Dziri et al.(2023)]{Dziri2023}
Dziri, N., et al. (2023). Faith and fate: Limits of transformers on compositionality. \emph{NeurIPS}.

\bibitem[GitClear(2025)]{GitClear2025}
GitClear. (2025). \emph{AI Copilot Code Quality: 2025 Research}. Available at: \url{https://www.gitclear.com/ai_assistant_code_quality_2025_research} (accessed 2026).

\bibitem[Huang et al.(2023)]{Huang2023}
Huang, J., et al. (2023). Large language models cannot self-correct reasoning yet. \emph{arXiv:2310.01798}.

\bibitem[Kahneman(2011)]{Kahneman2011}
Kahneman, D. (2011). \emph{Thinking, Fast and Slow}. Farrar, Straus and Giroux.

\bibitem[Karpathy(2025)]{Karpathy2025}
Karpathy, A. (2025). Vibe coding [Post]. \emph{X (formerly Twitter)}, February~2, 2025. \url{https://x.com/karpathy/status/1886192184808149383} (accessed 2026). The term was subsequently recognised as Collins English Dictionary Word of the Year 2025.

\bibitem[Kazman et al.(2000)]{Kazman2000}
Kazman, R., Klein, M., \& Clements, P. (2000). \emph{ATAM: Method for Architecture Evaluation}. Technical Report CMU/SEI-2000-TR-004, SEI/CMU.

\bibitem[Lehman(1980)]{Lehman1980}
Lehman, M. (1980). Programs, life cycles, and laws of software evolution. \emph{Proceedings of the IEEE}, 68(9), 1060--1076.

\bibitem[Leveson(2011)]{Leveson2011}
Leveson, N. (2011). \emph{Engineering a Safer World}. MIT Press.

\bibitem[Liu et al.(2023)]{Liu2023}
Liu, N.~F., et al. (2023). Lost in the middle: How language models use long contexts. \emph{arXiv:2307.03172}.

\bibitem[Martin(2003)]{Martin2003}
Martin, R. C. (2003). \emph{Agile Software Development: Principles, Patterns, and Practices}. Prentice Hall.

\bibitem[METR(2025)]{METR2025}
METR. (2025). Measuring the impact of early-2025 AI on experienced open-source developer productivity. \emph{arXiv:2507.09089}.

\bibitem[Meyer(1988)]{Meyer1988}
Meyer, B. (1988). \emph{Object-Oriented Software Construction}. Prentice Hall.

\bibitem[Meyer(1992)]{Meyer1992}
Meyer, B. (1992). Applying design by contract. \emph{Computer}, 25(10), 40--51.

\bibitem[Nygard(2011)]{Nygard2011}
Nygard, M. (2011). Documenting architecture decisions. \emph{Cognitect Blog}. Available at: \url{https://cognitect.com/blog/2011/11/15/documenting-architecture-decisions} (accessed 2026).

\bibitem[Lovable(2025)]{Lovable2025}
Lovable. (2025). Lovable reaches \$100M ARR. \emph{lovable.dev blog} (accessed 2026). Available at: \url{https://lovable.dev/blog/100m-arr}

\bibitem[Palmer(2025)]{Palmer2025}
Palmer, M. (2025). Statement on CVE-2025-48757: Lovable row level security vulnerability. \emph{mattpalmer.io}, May~29, 2025. \url{https://mattpalmer.io/posts/2025/05/statement-on-CVE-2025-48757/} (accessed 2026). Full technical disclosure at \url{https://mattpalmer.io/posts/2025/05/CVE-2025-48757/}. Official NVD entry: \url{https://nvd.nist.gov/vuln/detail/CVE-2025-48757}.

\bibitem[Perez et al.(2023)]{Perez2023}
Perez, E., Ringer, S., Lukošiūtė, K., Nguyen, K., Chen, E., Heiner, S., Pettit, C., Olsson, C., Kundu, S., Kadavath, S., Jones, A., Chen, A., Mann, B., Israel, B., Seethor, B., McKinnon, C., Maxwell, T., Telleen-Lawton, T., Hatfield-Dodds, Z., Kaplan, J., Clark, J., Brown, T., McCandlish, S., Askell, A., \& Ganguli, D. (2023). Discovering language model behaviors with model-written evaluations. \emph{arXiv:2212.09251}.

\bibitem[Y Combinator(2025)]{YC2025}
Y Combinator. (2025). YC Winter 2025 batch statistics. \emph{ycombinator.com} (accessed 2026). Available at: \url{https://www.ycombinator.com/blog/yc-stats-w25}

\bibitem[Popper(1959)]{Popper1959}
Popper, K. (1959). \emph{The Logic of Scientific Discovery}. Hutchinson.

\bibitem[Sharma et al.(2023)]{Sharma2023}
Sharma, M., Tong, M., Korbak, T., Duvenaud, D., Askell, A., Bowman, S.~R., Cheng, N., Durmus, E., Hatfield-Dodds, Z., Irving, G., Kravec, S., Maxwell, T., McCandlish, S., Ndousse, K., Rausch, O., Schiefer, N., Yan, D., Ziegler, D., \& Perez, E. (2023). Towards understanding sycophancy in language models. \emph{arXiv:2310.13548}.

\bibitem[Stack Overflow(2025)]{StackOverflow2025}
Stack Overflow. (2025). \emph{2025 Developer Survey}. Available at: \url{https://survey.stackoverflow.co/2025} (accessed 2026).

\bibitem[Ferrari et al.(2016)]{Ferrari2016tacit}
Ferrari, A., Spoletini, P., \& Gnesi, S. (2016). Ambiguity and tacit knowledge in requirements elicitation interviews. \emph{Requirements Engineering}, 21(3), 333--355. \url{https://doi.org/10.1007/s00766-016-0249-3}

\bibitem[Bano et al.(2019)]{Bano2019}
Bano, M., Zowghi, D., Ferrari, A., \& Spoletini, P. (2019). Teaching requirements elicitation interviews: an empirical study of learning from mistakes. \emph{Requirements Engineering}, 24(3), 259--289. \url{https://doi.org/10.1007/s00766-019-00313-0}

\bibitem[Hovsepyan et al.(2024)]{Hovsepyan2024}
Hovsepyan, A., et al. (2024). AutoSafeCoder: A multi-agent framework for securing LLM code generation through static analysis and fuzz testing. \emph{arXiv:2409.10737}.

\bibitem[Hasan et al.(2025)]{Hasan2025}
Hasan, M., et al. (2025). PREFACE: Property-driven reinforcement for automated code generation. \emph{Proceedings of the ACM/IEEE GLSVLSI 2025}.

\end{thebibliography}

\end{document}